\newcommand{\dmb}{\ensuremath{\dot{m_{\rm{B}}}}}
\newcommand{\dme}{\ensuremath{\dot{m_{\rm{E}}}}}

\newcommand{\ms}{MS 0735.6+7421}
\newcommand{\myi}{\ensuremath{I}}
\newcommand{\myv}{\ensuremath{V}}
\newcommand{\rbs}{RBS 797}
\newcommand{\reff}{\ensuremath{r_{\rm{eff}}}}
\newcommand{\rlos}{\ensuremath{r_{\rm{los}}}}
\newcommand{\mykeywords}{galaxies: active -- galaxies: clusters:
  general -- galaxies: clusters: intracluster medium -- X-rays:
  galaxies: clusters}
\newcommand{\mystitle}{\rbs\ AGN Outburst}
\newcommand{\mytitle}{A Powerful AGN Outburst in RBS 797}

\documentclass[iop]{emulateapj}
\usepackage{apjfonts,common,graphicx}
\usepackage[pagebackref,
  pdftitle={\mytitle},
  pdfauthor={K. W. Cavagnolo},
  pdfkeywords={\mykeywords},
  pdfsubject={Astrophysical Journal},
  pdfproducer={ps2pdf},
  pdfcreator={LaTeX with hyperref}
  pdfdisplaydoctitle=true,
  colorlinks=true,
  citecolor=blue,
  linkcolor=blue,
  urlcolor=blue]{hyperref}
\slugcomment{Accepted to ApJ}
\begin{document}
\title{\mytitle}
\shorttitle{\mystitle}
\author{
  K. W. Cavagnolo,\altaffilmark{1}
  B. R. McNamara,\altaffilmark{2,3,4}
  M. W. Wise,\altaffilmark{5}
  P. E. J. Nulsen,\altaffilmark{4}\\
  M. Br\"uggen,\altaffilmark{6}
  M. Gitti,\altaffilmark{4,7} and
  D. A. Rafferty\altaffilmark{8}
}
\shortauthors{Cavagnolo et al.}
\email{cavagnolo@oca.eu}
\altaffiltext{1}{UNS, CNRS UMR 6202 Cassiop\'{e}e, Observatoire de la
  C\^{o}te d'Azur, Nice, France.}
\altaffiltext{2}{University of Waterloo, Department of Physics \&
  Astronomy, Waterloo, Canada.}
\altaffiltext{3}{Perimeter Institute for Theoretical Physics,
  Waterloo, Canada.}
\altaffiltext{4}{Harvard-Smithsonian Center for Astrophysics,
  Cambridge, USA.}
\altaffiltext{5}{University of Amsterdam, Astronomical Institute Anton
  Pannekoek, Amsterdam, Netherlands.}
\altaffiltext{6}{Jacobs University Bremen, Bremen, Germany.}
\altaffiltext{7}{Astronomical Observatory of Bologna-INAF, Bologna,
  Italy.}
\altaffiltext{8}{Leiden Observatory, University of Leiden, Leiden,
  Netherlands.}

\begin{abstract}
  Utilizing $\sim 50$ ks of Chandra X-ray Observatory imaging, we
  present an analysis of the intracluster medium (ICM) and cavity
  system in the galaxy cluster \rbs. In addition to the two previously
  known cavities in the cluster core, the new and deeper X-ray image
  has revealed additional structure associated with the active
  galactic nucleus (AGN). The surface brightness decrements of the two
  cavities are unusually large, and are consistent with elongated
  cavities lying close to our line-of-sight. We estimate a total AGN
  outburst energy and mean jet power of $\approx 3 \dash 6 \times
  10^{60}$ erg and $\approx 3 \dash 6 \times 10^{45} ~\lum$,
  respectively, depending on the assumed geometrical configuration of
  the cavities. Thus, \rbs\ is apparently among the the most powerful
  AGN outbursts known in a cluster. The average mass accretion rate
  needed to power the AGN by accretion alone is $\sim 1 ~\msolpy$. We
  show that accretion of cold gas onto the AGN at this level is
  plausible, but that Bondi accretion of the hot atmosphere is
  probably not. The BCG harbors an unresolved, non-thermal nuclear
  X-ray source with a bolometric luminosity of $\approx 2 \times
  10^{44} ~\lum$. The nuclear emission is probably associated with a
  rapidly-accreting, radiatively inefficient accretion flow. We
  present tentative evidence that star formation in the BCG is being
  triggered by the radio jets and suggest that the cavities may be
  driving weak shocks ($M \sim 1.5$) into the ICM, similar to the
  process in the galaxy cluster \ms.
\end{abstract}


\section{Introduction}
\label{sec:intro}

Evidence gathered over the last decade suggests that the growth of
galaxies and supermassive black holes (SMBHs) are coupled, and that
energetic feedback from active galactic nuclei (AGN) strongly
influences galaxy evolution \citep[\eg][]{1995ARA&A..33..581K,
  1998A&A...331L...1S, 2000MNRAS.311..576K, 2002ApJ...574..740T}. The
discovery of AGN induced cavities in the hot halos surrounding many
massive galaxies has strengthened this idea \citep[see][for
  reviews]{cfreview, mcnamrev} by revealing that AGN mechanical
heating is capable of regulating halo radiative cooling
\citep[\eg][]{birzan04, dunn06, rafferty06}. Current models of
radio-mode AGN feedback posit that cooling processes in a galaxy's hot
halo drives mass accretion onto a central SMBH, promoting AGN activity
that eventually offsets halo cooling via a thermally regulated
feedback loop \citep[\eg][]{croton06, bower06, sijacki07}. While there
is direct evidence that halo cooling and feedback are linked
\citep[\eg][]{haradent, rafferty08}, the observational constraints on
how AGN are fueled and powered are more difficult to establish.

Gas accretion alone can, in principle, fuel most AGN
\citep[\eg][]{pizzolato05, 2006MNRAS.372...21A}. However, for some
relatively gas-poor systems hosting energetic AGN, for example,
Hercules A, Hydra A, \ms, and 3C 444 where the output exceeds
$10^{61}$ erg \citep{herca, hydraa, ms0735, 2010arXiv1011.6405C}, it
appears that gas accretion alone may have difficulty sustaining their
AGN unless the accretion is unusually efficient. This problem has lead
to speculation that some BCGs may host ultramassive black holes
\citep[$> 10^{10} ~\msol$; \eg][]{msspin}, or that some AGN are
powered by the release of energy stored in a rapidly-spinning SMBH
\citep[\eg][]{minaspin}. In this paper, we explore these and other
issues through an analysis of the powerful AGN outburst in the
\object{RBS 797} cluster.

The discovery of AGN-induced cavities in the intracluster medium (ICM)
of the galaxy cluster \rbs\ was first reported by \citet{schindler01}
using data from the Chandra X-ray Observatory (\cxo). Multifrequency
radio observations showed that the cavities are co-spatial with
extended radio emission centered on a strong, jetted radio source
coincident with the \rbs\ brightest cluster galaxy (BCG)
\citep{2002astro.ph..1349D, gitti06, birzan08}. The observations
implicate an AGN in the BCG as the cavities' progenitor, and
\citet[][hereafter B04]{birzan04} estimate the AGN deposited $\approx
10^{60}$ erg of energy into the ICM at a rate of $\approx 10^{45}
~\lum$.

The B04 analysis assumed the cavities are roughly spherical, symmetric
about the plane of the sky, and that their centers lie in a plane
passing through the central AGN and perpendicular to our
line-of-sight. However, the abnormally deep surface brightness
decrements of the cavities, and the nebulous correlation between the
radio and X-ray morphologies, suggests the system may be more complex
than B04 assumed. Using a longer, follow-up \cxo\ observation, we
conclude that the cavities are probably elongated along the
line-of-sight and provide evidence that additional cavities may be
present at larger radii. We conservatively estimate the total AGN
energy output and power are six times larger than the B04 values,
$\approx 6 \times 10^{60}$ erg and $\approx 6 \times 10^{45} ~\lum$
respectively.

Reduction of the X-ray and radio data is discussed in Section
\ref{sec:obs}. Interpretation of observational results are given
throughout Section \ref{sec:results}, and a brief summary concludes
the paper in Section \ref{sec:con}. \LCDM. At a redshift of $z =
0.354$, the look-back time is 3.9 Gyr, $\da = 4.996$ kpc
arcsec$^{-1}$, and $\dl = 1889$ Mpc. All spectral analysis errors are
90\% confidence, while all other errors are 68\% confidence.

\section{Observations}
\label{sec:obs}

\subsection{X-ray Data}
\label{sec:xray}

\begin{figure}
  \begin{center}
    \begin{minipage}{\linewidth}
      \includegraphics*[width=\textwidth, trim=0mm 0mm 0mm 0mm, clip]{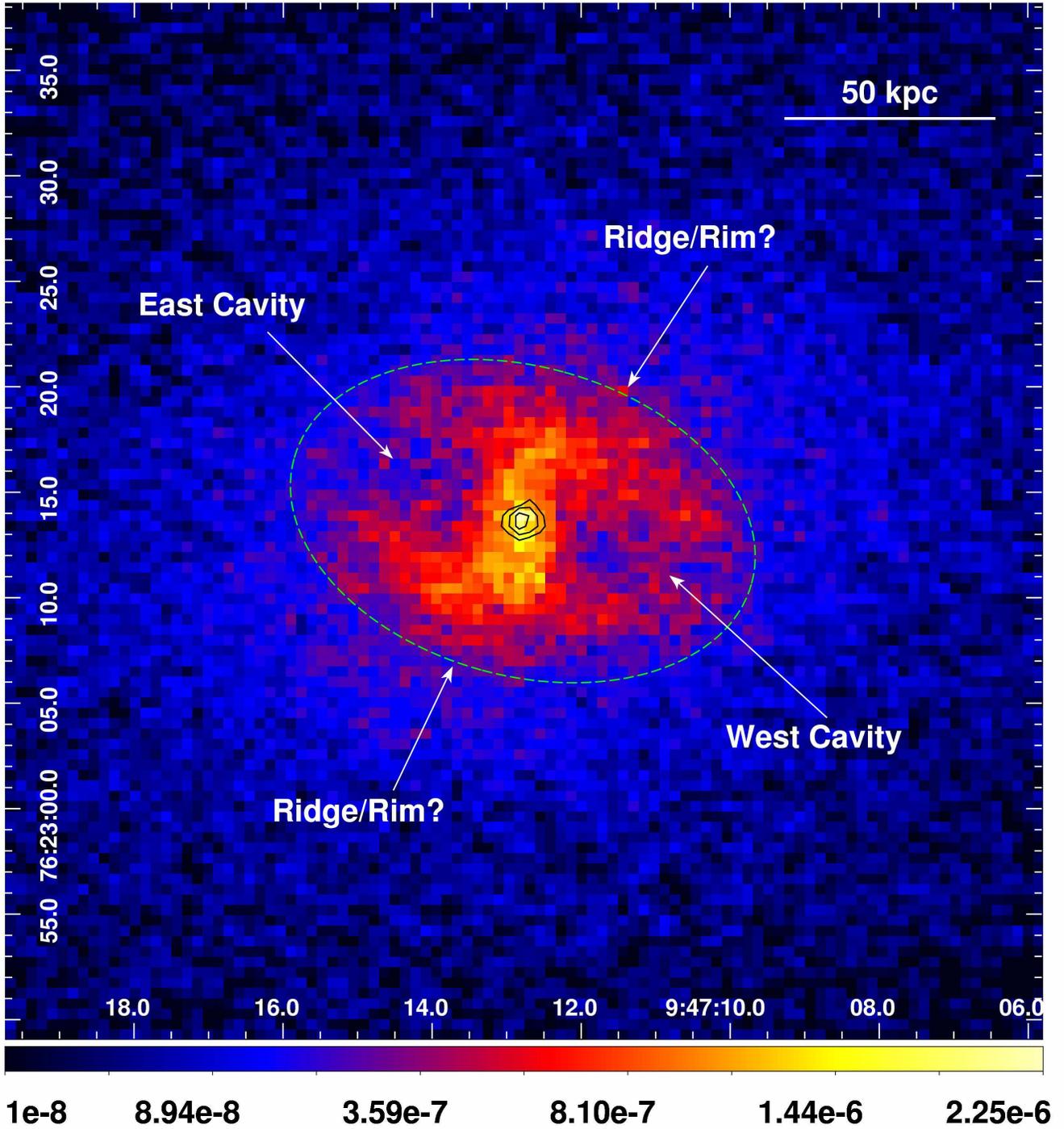}
    \end{minipage}
    \caption{Fluxed, unsmoothed 0.7--2.0 keV clean image of \rbs\ in
      units of ph \pcmsq\ \ps\ pix$^{-1}$. Image is $\approx 250$ kpc
      on a side and coordinates are J2000 epoch. Black contours in the
      nucleus are 2.5--9.0 keV X-ray emission of the nuclear point
      source; the outer contour approximately traces the 90\% enclosed
      energy fraction (EEF) of the \cxo\ point spread function. The
      dashed green ellipse is centered on the nuclear point source,
      encloses both cavities, and was drawn by-eye to pass through the
      X-ray ridge/rims.}
    \label{fig:img}
  \end{center}
\end{figure}

\rbs\ was observed with \cxo\ in October 2000 for 11.4 ks using the
ACIS-I array (\dataset [ADS/Sa.CXO#Obs/02202] {ObsID 2202}) and in
July 2007 for 38.3 ks using the ACIS-S array (\dataset
[ADS/Sa.CXO#Obs/07902] {ObsID 7902}). Datasets were reduced using
\ciao\ and \caldb\ versions 4.2. Events were screened of cosmic rays
using \asca\ grades and {\textsc{vfaint}} filtering. The level-1 event
files were reprocessed to apply the most up-to-date corrections for
the time-dependent gain change, charge transfer inefficiency, and
degraded quantum efficiency of the ACIS detector. The afterglow and
dead area corrections were also applied. Time intervals affected by
background flares exceeding 20\% of the mean background count rate
were isolated and removed using light curve filtering. The final,
combined exposure time is 48.8 ks. Point sources were identified and
removed via visual inspection and use of the \ciao\ tool
{\textsc{wavdetect}}. We refer to the \cxo\ data free of point sources
and flares as the ``clean'' data. A mosaiced, fluxed image (see Figure
\ref{fig:img}) was generated by exposure correcting each clean dataset
and reprojecting the normalized images to a common tangent point.

\subsection{Radio Data}
\label{sec:radio}

Very Large Array (\vla) radio images at 325 MHz (A-array), 1.4 GHz (A-
and B-array), 4.8 GHz (A-array), and 8.4 GHz (D-array) are presented
in \citet{gitti06} and \citet{birzan08}. Our re-analysis of the
archival \vla\ radio observations yielded no significant differences
with these prior studies. Using the rms noise ($\sigma_{\rm{rms}}$)
values given in \citet{gitti06} and \citet{birzan08} for each
observation, emission contours between $3\sigma_{\rm{rms}}$ and the
peak image intensity were generated. These are the contours referenced
and shown in all subsequent discussion and figures.

\section{Results}
\label{sec:results}

\subsection{Cavity Morphologies and ICM Substructure}
\label{sec:morph}

Shown in Figure \ref{fig:img} is the 0.7--2.0 keV \cxo\ mosaiced clean
image. Outside of $\approx 50$ kpc, the global ICM morphology is
regular and elliptical in shape, with the appearance of being
elongated along the northwest-southeast direction. The cavities
discovered by \citet{schindler01} are clearly seen in the cluster core
east and west of the nuclear X-ray source and appear to be enclosed by
a thick, bright, elliptical ridge of emission which we discuss further
in Section \ref{sec:ecav}. The western cavity has more internal
structure, and its boundaries are less well-defined, than the eastern
cavity. The emission from the innermost region of the core is
elongated north-south and has a distinct `S'-shape punctuated by a
hard nuclear X-ray source.

\begin{figure}
  \begin{center}
    \begin{minipage}{0.495\linewidth}
      \includegraphics*[width=\textwidth, trim=0mm 0mm 0mm 0mm, clip]{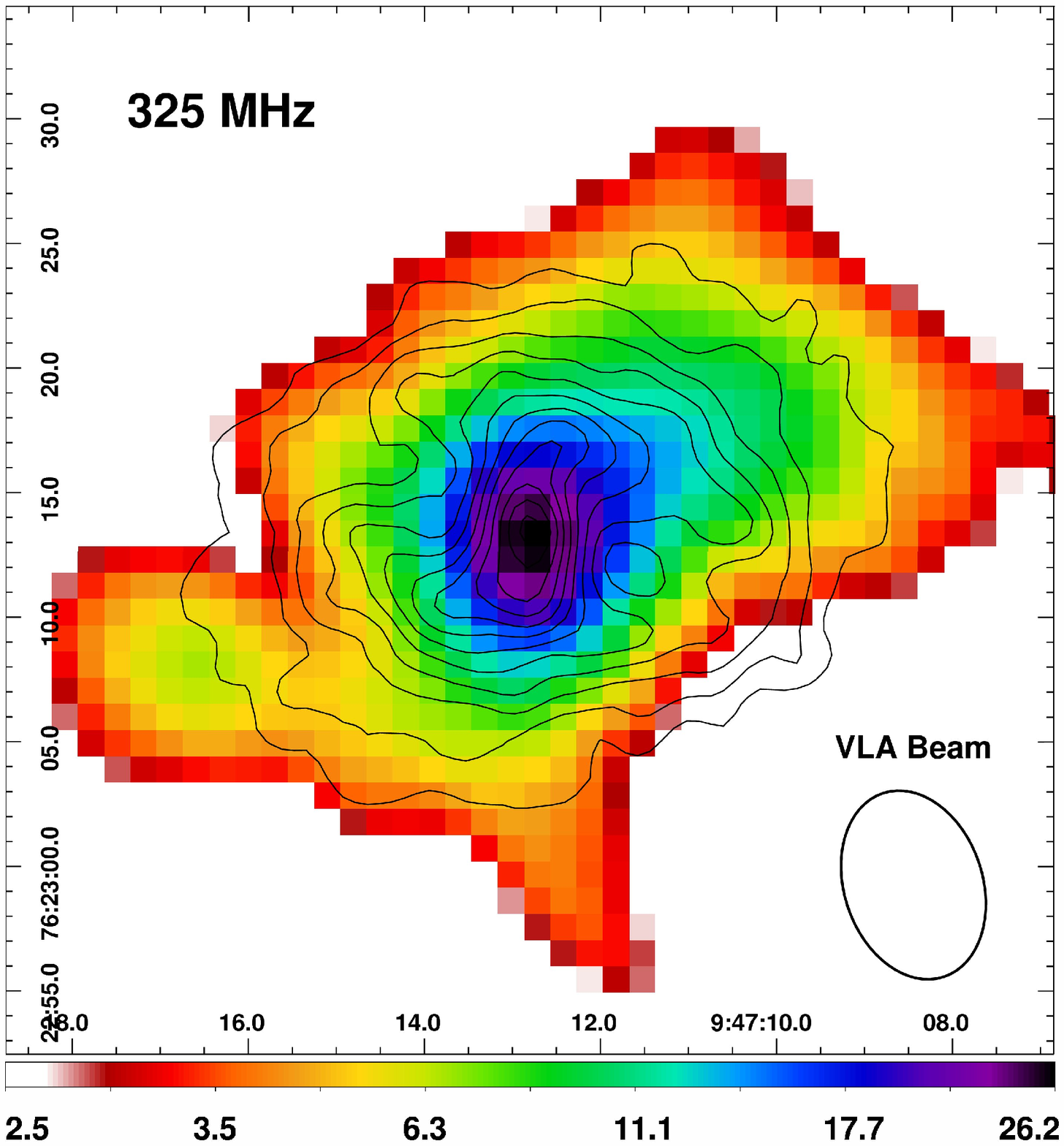}
    \end{minipage}
   \begin{minipage}{0.495\linewidth}
      \includegraphics*[width=\textwidth, trim=0mm 0mm 0mm 0mm, clip]{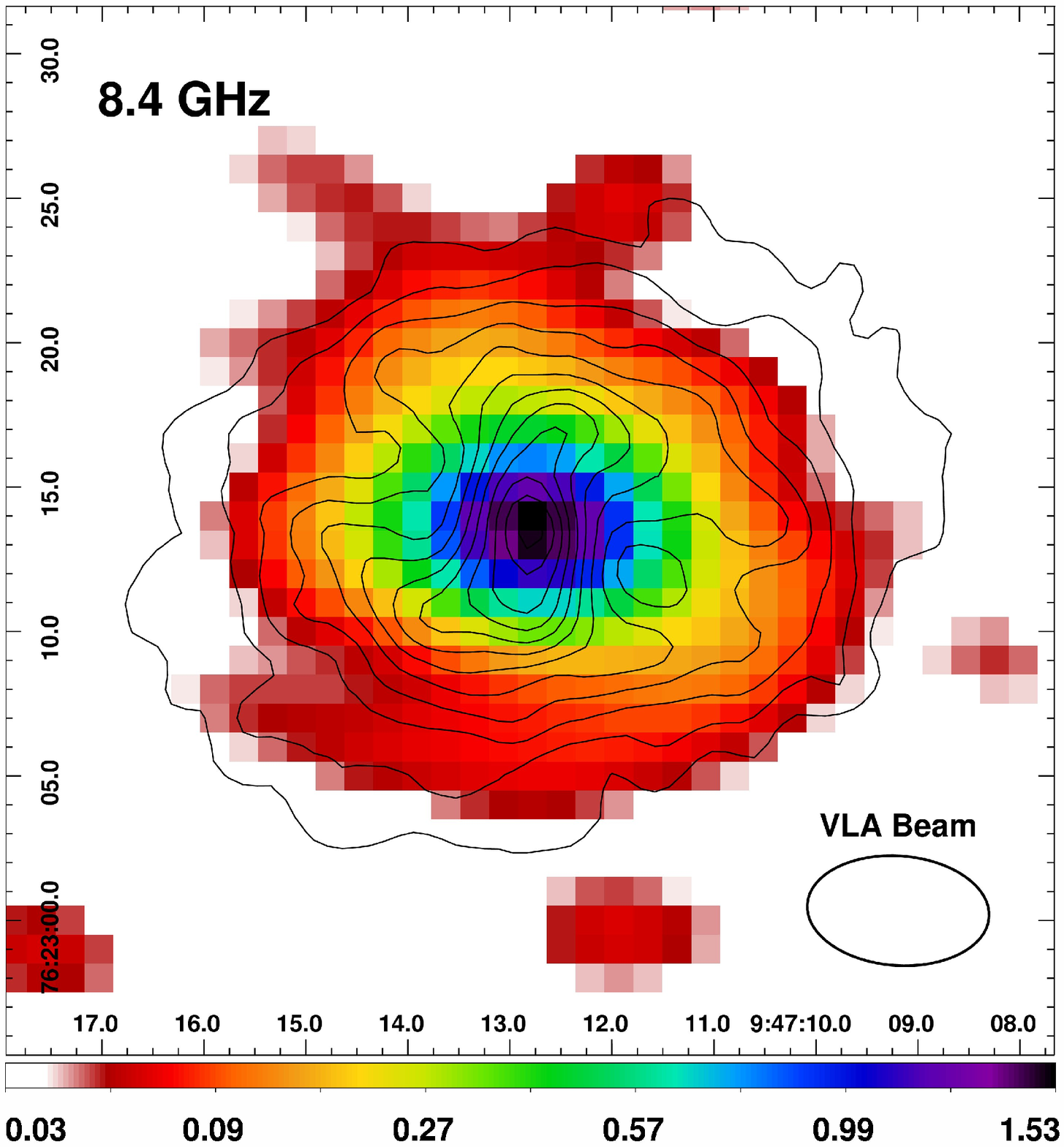}
   \end{minipage}
   \begin{minipage}{0.495\linewidth}
      \includegraphics*[width=\textwidth, trim=0mm 0mm 0mm 0mm, clip]{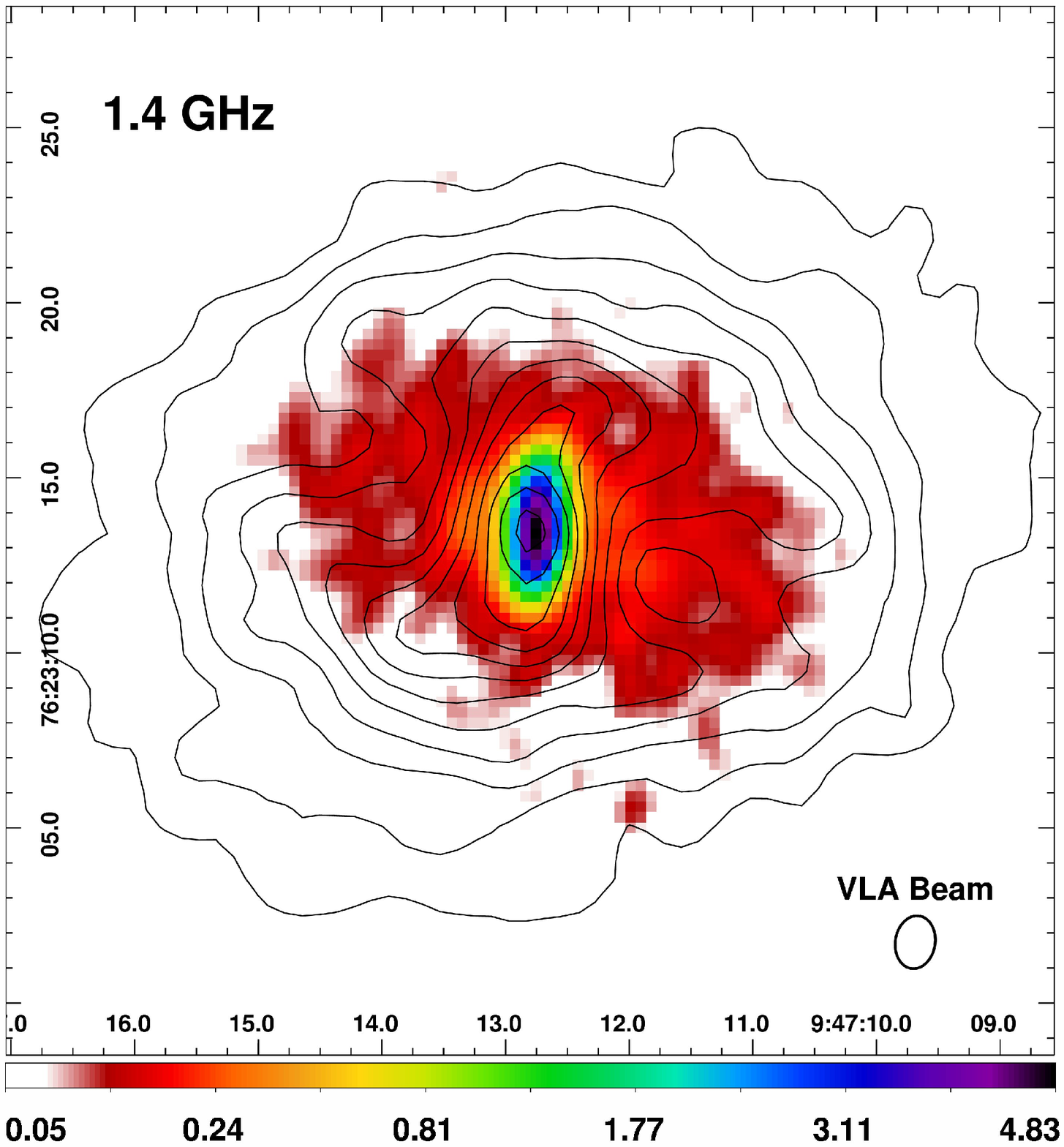}
    \end{minipage}
    \begin{minipage}{0.495\linewidth}
      \includegraphics*[width=\textwidth, trim=0mm 0mm 0mm 0mm, clip]{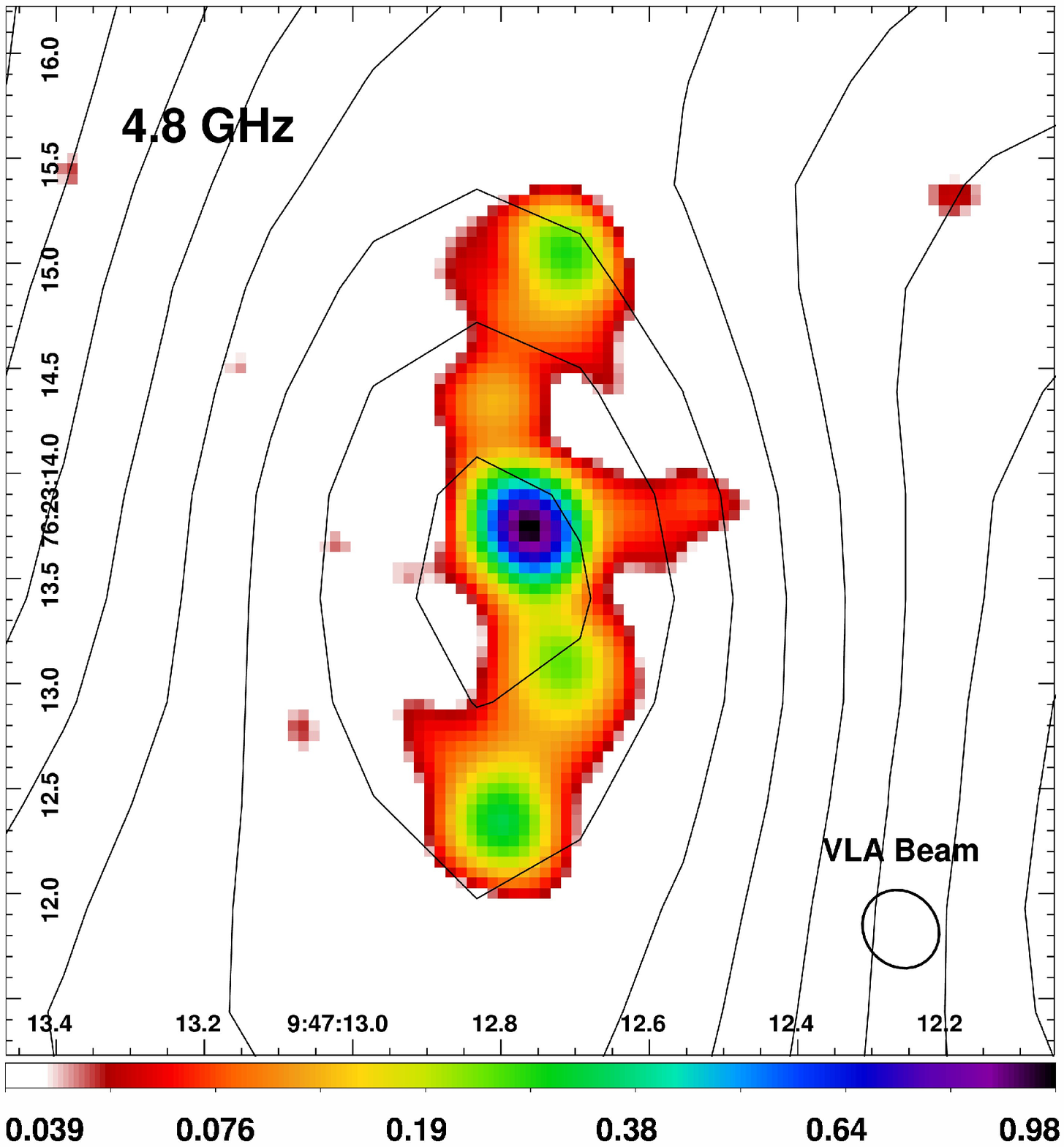}
    \end{minipage}
     \caption{Radio images of \rbs\ overlaid with black contours
       tracing ICM X-ray emission. Images are in mJy beam$^{-1}$ with
       intensity beginning at $3\sigma_{\rm{rms}}$ and ending at the
       peak flux, and are arranged by decreasing size of the
       significant, projected radio structure. X-ray contours are from
       $2.3 \times 10^{-6}$ to $1.3 \times 10^{-7}$ ph
       \pcmsq\ \ps\ pix$^{-1}$ in 12 square-root steps. {\it{Clockwise
           from top left}}: 325 MHz \vla\ A-array, 8.4 GHz
       \vla\ D-array, 4.8 GHz \vla\ A-array, and 1.4 GHz
       \vla\ A-array.}
    \label{fig:composite}
  \end{center}
\end{figure}

Multifrequency radio images overlaid with ICM X-ray contours are shown
in Figure \ref{fig:composite}. \rbs\ radio properties are discussed in
\citet{gitti06}, and we summarize here. As seen in projection, the
nuclear 4.8 GHz jets are almost orthogonal to the axis connecting the
cavities. Radio imaging at $\approx 5\arcs$ resolution reveals the 325
MHz, 1.4 GHz, and 8.4 GHz radio emission are diffuse and extend
well-beyond the cavities, more similar to the morphology of a radio
mini-halo than relativistic plasma confined to the cavities
\citep[][Doria et al., in preparation]{2008A&A...486L..31C}. Of all
the radio images, the 1.4 GHz emission imaged at $\approx 1 \arcs$
resolution most closely traces the cavity morphologies, yet it is
still diffuse and uniform over the cavities with little structure
outside the radio core. Typically, the connection between a cavity
system, coincident radio emission, and the progenitor AGN is
unambiguous. This is not the case for \rbs, which suggests the cavity
system may be more complex than it appears. Indeed, there is a hint of
barely resolved structure in the Chandra image associated with the 4.8
GHz radio structure, with X-ray deficits just beyond the tips of the
radio jets, bounded to one side by the 'S' shaped feature noted
above. This is suggestive of potentially small cavities from a new
outburst episode with asymmetric bright rims created by young radio
jets.

To better reveal the cavity morphologies, residual X-ray images of
\rbs\ were constructed by modeling the ICM emission and subtracting it
off. The X-ray isophotes of two exposure-corrected images -- one
smoothed by a $1\arcs$ Gaussian and another by a $3\arcs$ Gaussian --
were fitted with ellipses using the \iraf\ task \textsc{ellipse}. The
ellipse centers were fixed at the location of the BCG X-ray point
source, and the eccentricities and position angles were free to
vary. A 2D surface brightness model was created from each fit using
the \iraf\ task \textsc{bmodel}, normalized to the parent X-ray image,
and then subtracted off. The residual images are shown in Figure
\ref{fig:subxray}.

In addition to the central east and west cavities (labeled E1 and W1),
tentative evidence for depressions north and south of the nucleus
(labeled N1 and S1) are revealed. N1 and S1 lie along the 4.8 GHz jet
axis and are coincident with spurs of significant 1.4 GHz emission. A
possible depression coincident with the southeastern concentration of
325 MHz emission is also found (labeled E2), but no radio counterpart
on the opposite side of the cluster is seen (labeled W2). The N1, S1,
and E2 depressions do not show-up in surface brightness profiles
extracted in wedges passing through each feature. So while coincidence
with the radio emission hints at additional cavities, they may be
spurious structures and a deeper image is required to confirm
them. There is also an X-ray edge which extends southeast from E2 and
sits along a ridge of 325 MHz and 8.4 GHz emission. No substructure
associated with the western-most knot of 325 MHz emission is found,
but there is a stellar object co-spatial with this region. The X-ray
and radio properties of the object are consistent with those of a
galactic RS CVn star \citep{1993RPPh...56.1145S} -- if the star is
less than 1 kpc away, $\lx \la 10^{31} ~\lum$ and $L_{325} \sim
10^{27} ~\lum$ -- suggesting the western 325 MHz emission may not be
associated with the cluster.

\subsection{Radial ICM Properties}
\label{sec:icm}

\begin{figure}
  \begin{center}
    \begin{minipage}{\linewidth}
      \includegraphics*[width=\textwidth]{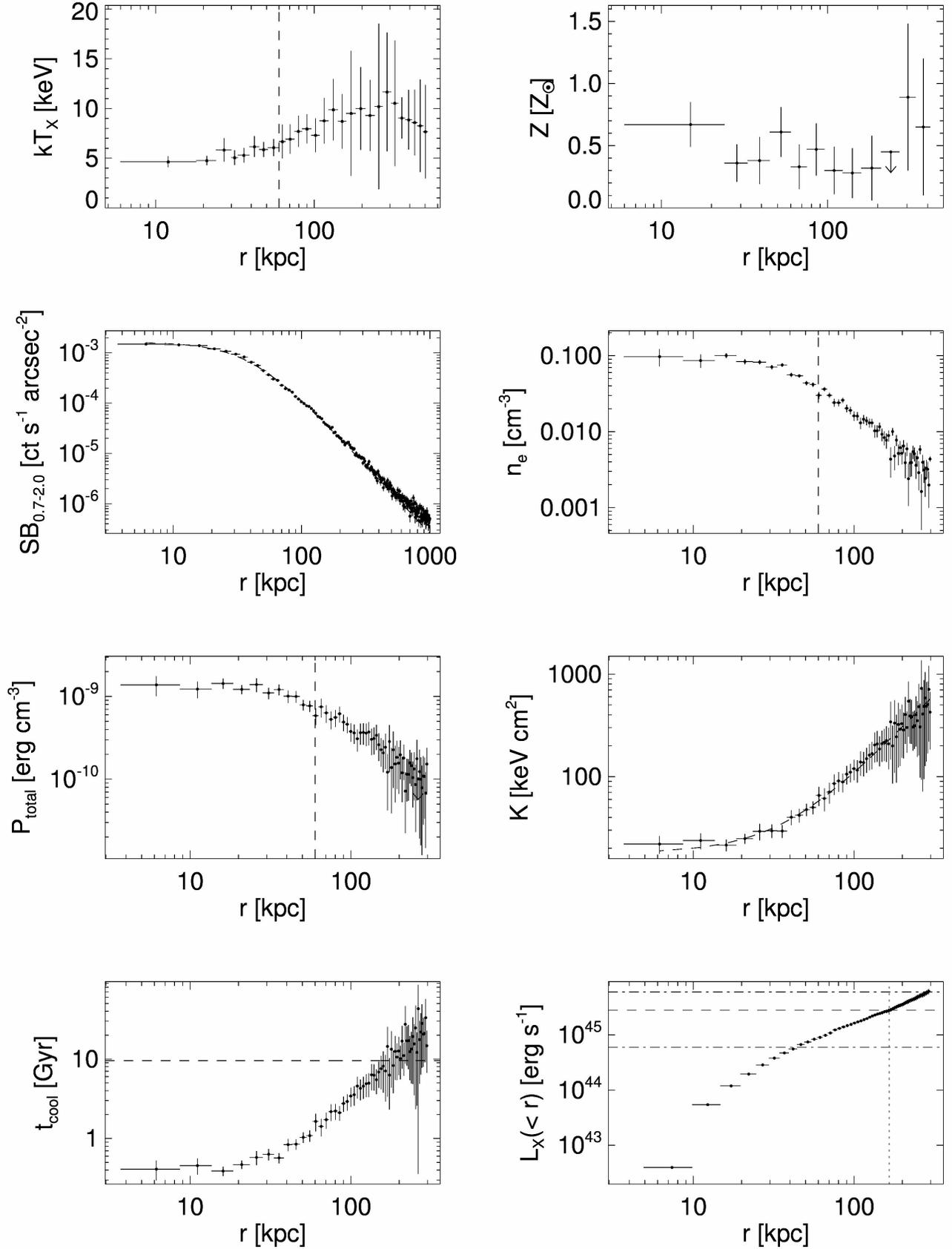}
      \caption{Gallery of radial ICM profiles. Vertical black dashed
        lines mark the approximate end-points of both
        cavities. Horizontal dashed line on cooling time profile marks
        age of the Universe at redshift of \rbs. For X-ray luminosity
        profile, dashed line marks \lcool, and dashed-dotted line
        marks \pcav.}
      \label{fig:gallery}
    \end{minipage}
  \end{center}
\end{figure}

In order to analyze the \rbs\ cavity system and AGN outburst
energetics in-detail, the radial ICM density, temperature, and
pressure structure need to be measured. The radial profiles are shown
in Figure \ref{fig:gallery}. A temperature (\tx) profile was created
by extracting spectra from concentric circular annuli (2500 source
counts per annulus) centered on the cluster X-ray peak, binning the
spectra to 25 counts per energy channel, and then fitting each
spectrum in \xspec\ 12.4 \citep{xspec}. Spectra were modeled with an
absorbed, single-component \mekal\ model \citep{mekal1} over the
energy range 0.7--7.0 \keV. For each annulus, weighted responses were
created and a background spectrum was extracted from the ObsID matched
\caldb\ blank-sky dataset normalized using the ratio of 9--12 keV
count rates for an identical off-axis, source-free region of the
blank-sky and target datasets. The absorbing Galactic column density
was fixed to $\nhgal = 2.28 \times 10^{20} ~\pcmsq$ \citep{lab}, and a
spectral model for the Galactic foreground was included as an
additional, fixed background component during spectral fitting (see
\citealt{2005ApJ...628..655V} and \citealt{xrayband} for method). Gas
metal abundance was free to vary and normalized to the \citet{ag89}
solar ratios. Spectral deprojection using the {\textsc{projct}}
\xspec\ model did not produce significantly different results, thus
only projected quantities are discussed.

A 0.7--2.0 keV surface brightness profile was extracted from the
\cxo\ mosaiced clean image using concentric $1\arcs$ wide elliptical
annuli centered on the BCG X-ray point source (central $\approx
1\arcs$ and cavities excluded). A deprojected electron density
(\nelec) profile was derived from the surface brightness profile using
the method of \citet{kriss83} which incorporates the 0.7--2.0 keV
count rates and best-fit normalizations from the spectral analysis
\citep[see][for details]{accept}. Errors for the density profile were
estimated using 5000 Monte Carlo simulations of the original surface
brightness profile. A total gas pressure profile was calculated as $P
= n \tx$ where $n \approx 2.3 \nH$ and $\nH \approx
\nelec/1.2$. Profiles of enclosed X-ray luminosity, entropy ($K = \tx
\nelec^{-2/3}$) and cooling time ($\tcool = 3 n \tx~(2 \nelec \nH
\Lambda)^{-1}$, where $\Lambda$ is the cooling function), were also
generated. Errors for each profile were determined by summing the
individual parameter uncertainties in quadrature.

The radial profiles are consistent with those of a normal cool core
cluster: outwardly increasing temperature profile, centrally peaked
abundance profile, central cooling time of $\approx 400$ Myr within 9
kpc of the cluster center, and a pronounced low entropy core. Fitting
the ICM entropy profile with the function $K = \kna +\khun (r/100
~\kpc)^{\alpha}$, where \kna\ [\ent] is core entropy, \khun\ [\ent] is
a normalization at 100 kpc, and $\alpha$ is a dimensionless index,
reveals $\kna = 17.9 \pm 2.2 ~\ent$, $\khun = 92.1 \pm 6.2 ~\ent$, and
$\alpha = 1.65 \pm 0.11$ for \chisq(DOF)$ = 9.3(57)$. Further, the
radial analysis does not indicate any significant temperature,
density, or pressure discontinuities to signal the existence of
large-scale shocks or cold fronts.

\subsection{ICM Cavities}
\label{sec:cavities}

\subsubsection{Cavity Energetics}
\label{sec:ecav}

\begin{figure}
  \begin{center}
    \begin{minipage}{\linewidth}
      \setlength\fboxsep{0pt}
      \setlength\fboxrule{0.5pt}
      \fbox{\includegraphics*[width=\textwidth]{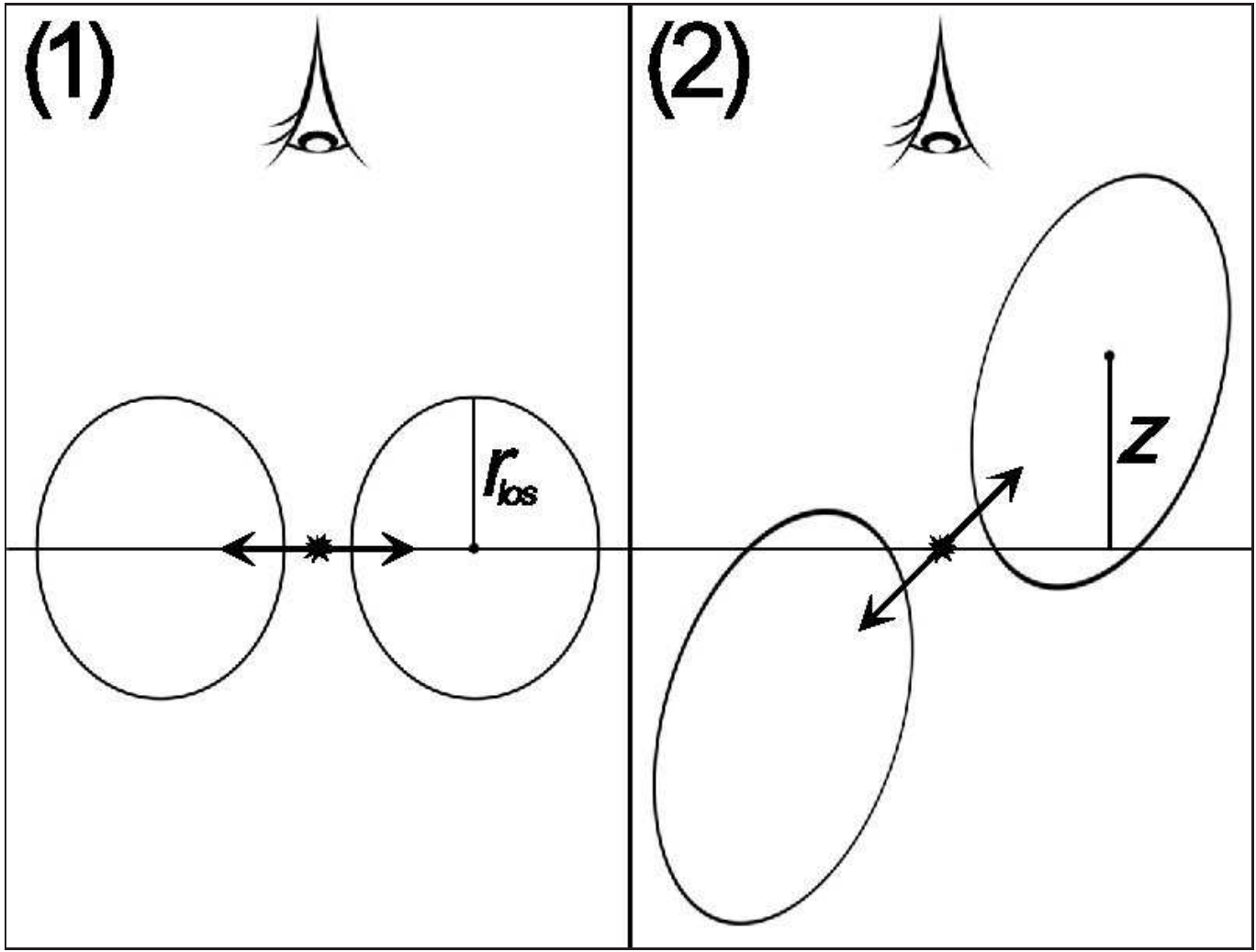}}
    \end{minipage}
    \caption{Cartoon of possible cavity configurations. Arrows denote
      direction of AGN outflow, ellipses outline cavities, \rlos\ is
      line-of-sight cavity depth, and $z$ is the height of a cavity's
      center above the plane of the sky. {\it{Left:}} Cavities which
      are symmetric about the plane of the sky, have $z=0$, and are
      inflating perpendicular to the line-of-sight. {\it{Right:}}
      Cavities which are larger than left panel, have non-zero $z$,
      and are inflating along an axis close to our line-of-sight.}
    \label{fig:config}
  \end{center}
\end{figure}

Of all the ICM substructure, the E1 and W1 cavities are unambiguous
detections, and their energetics were determined following standard
methods (see B04). Our fiducial cavity configuration assumes the
cavities are symmetric about the plane of the sky, and the cavity
centers lie in a plane which is perpendicular to our line-of-sight and
passes through the central AGN (configuration-1 in Figure
\ref{fig:config}). Hereafter, we denote the line-of-sight distance of
a cavity's center from this plane as $z$, and the cavity radius along
the line-of-sight as \rlos. The volume of a cavity with $z = 0$ is
thus $V = 4\pi a b \rlos/3$ where $a$ and $b$ are the projected
semi-major and -minor axes, respectively, of the by-eye determined
ellipses in Figure \ref{fig:subxray}. The projected morphologies of E1
and W1 are similar enough that congruent regions were used for
both. Initially, the cavities were assumed to be roughly spherical and
\rlos\ was set equal to the projected effective radius, $\reff =
\sqrt{ab}$. For configuration-1, the distance of each cavity from the
central AGN, $D$, is simply the projected distance from the ellipse
centers to the BCG X-ray point source. A systematic error of 10\% is
assigned to the cavity dimensions.

Cavity ages were estimated using the \tsonic, \tbuoy, and
\trefill\ timescales discussed in B04. The \tsonic\ age assumes a
cavity reaches its present distance from the central AGN by moving at
the local sound speed, while \tbuoy\ assumes the cavity buoyantly
rises at its terminal velocity, and \trefill\ is simply the time
required to refill the volume displaced by the cavity. The energy
required to create each cavity is estimated by its enthalpy, $\ecav =
\gamma PV/(\gamma-1)$, which was calculated assuming cavity pressure
support is provided by a relativistic plasma ($\gamma = 4/3$). We
assume the cavities are buoyant structures and that the time-averaged
power needed to create each cavity is $\pcav = \ecav/\tbuoy$. The
individual cavity properties are given in Table \ref{tab:cavities} and
the aggregate properties in Table \ref{tab:totals}.

For the representative assumption of configuration-1, we measure a
total cavity energy $\ecav = 3.23 ~(\pm 1.16) \times 10^{60}$ erg and
total power $\pcav = 3.34 ~(\pm 1.41) \times 10^{45} ~\lum$. These
values are larger than those of B04 ($\ecav = 1.52 \times 10^{60}$
erg, $\pcav = 1.13 \times 10^{44} ~\lum$) by a factor of $\approx 3$
due to our larger E1 and W1 volumes. Using the B04 cavity volumes in
place of our own produces no significant differences between our
energetics calculations and those of B04. The largest uncertainties in
the energetic calculations are the cavity volumes and their
3-dimensional locations in the ICM, issues we consider next.

\subsubsection{Cavity Surface Brightness Decrements}
\label{sec:dec}

A cavity's morphology and location in the ICM affects the X-ray
surface brightness decrement it induces. If the surface brightness of
the undisturbed ICM can be estimated, then the decrement is useful for
constraining the cavity line-of-sight size and, through simple
geometric calculation, its distance from the launch point
\citep[see][for details]{hydraa}. We adopt the \citet{hydraa}
definition of surface brightness decrement, $y$, as the ratio of the
X-ray surface brightness inside the cavities to the value of the
best-fit ICM surface brightness $\beta$-model at the same radius. Note
the potential for confusion, since $y=1$ if there is no decrement and
while small values of $y$ correspond to large decrements. Consistent
with the analysis of \citet{schindler01}, we find the best-fit
$\beta$-model has parameters $S_0 = 1.65 ~(\pm 0.15) \times 10^{-3}$
\sbr, $\beta = 0.62 \pm 0.04$, and $r_{\rm{core}} = 7.98\arcs \pm
0.08$ for \chisq(DOF) = 79(97). Using a circular aperture with radius
$1\arcs$ centered on the deepest part of each cavity, we measure mean
decrements of $\bar{y}_{\rm{W1}} = 0.50 \pm 0.18$ and
$\bar{y}_{\rm{E1}} = 0.52 \pm 0.23$, with minima of
$y^{\rm{min}}_{\rm{W1}} = 0.44$ and $y^{\rm{min}}_{\rm{E1}} = 0.47$.

To check if the representative approximation for cavity
configuration-1 can produce the measured decrements, the best-fit
$\beta$-model was integrated over each cavity with a column of gas
equal to $2\rlos$ excluded. For this case, the most significant
decrement obtained was $y = 0.67$, indicating that $r_{los}$ must
exceed $r_{eff}$. If the centers of E1 and W1 lie in the plane of the
sky, the minimum line-of-sight depths needed to reproduce the
decrements of E1 and W1 are $\rlos^{\rm{E1}} = 23.4$ kpc and
$\rlos^{\rm{W1}} = 25.9$ kpc. The energetics for these morphologies
are given as configuration-1a in Tables \ref{tab:cavities} and
\ref{tab:totals}. If the cavities are moving away from the AGN in the
plane of the sky, then it is surprising to find that each has its
shortest axis in the plane of the sky and perpendicular to its
direction of motion. However, if the bright rims are masking the true
extent of the cavities along this short axis, the cavities may be
larger than we realize.

\begin{figure}
  \begin{center}
    \begin{minipage}{0.495\linewidth}
      \includegraphics*[width=\textwidth, trim=25mm 0mm 40mm 10mm, clip]{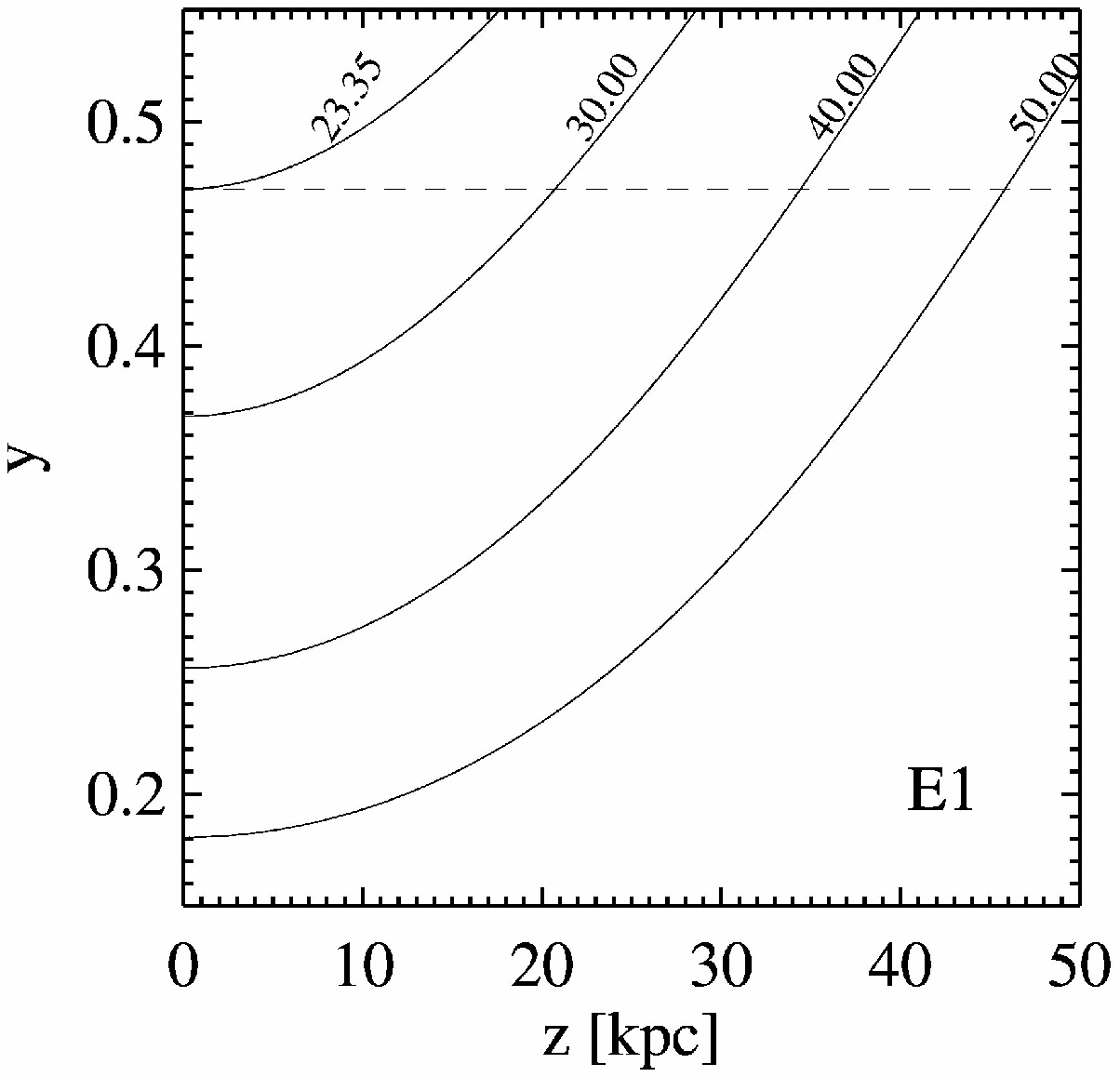}
    \end{minipage}
    \begin{minipage}{0.495\linewidth}
      \includegraphics*[width=\textwidth, trim=25mm 0mm 40mm 10mm, clip]{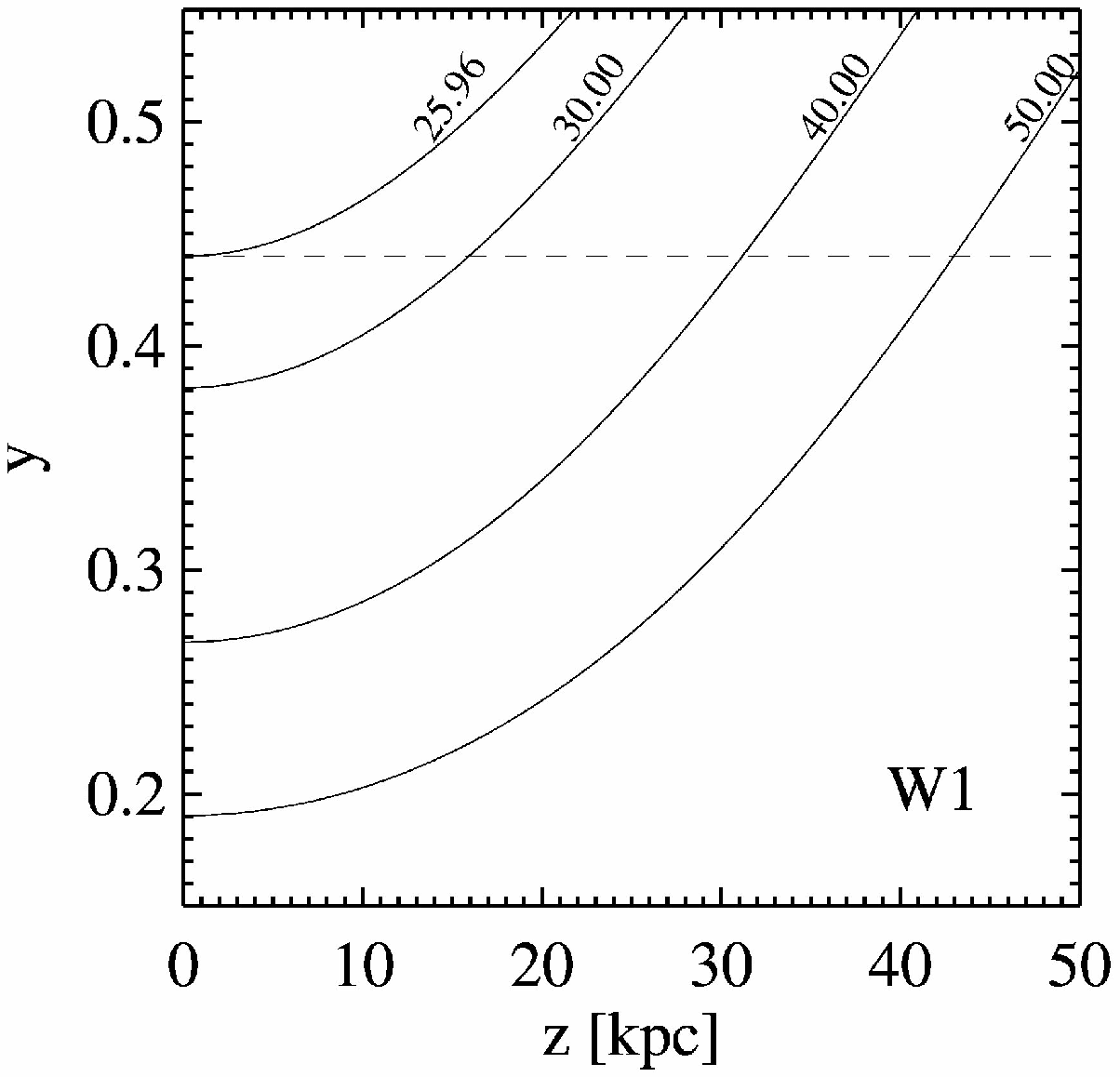}
    \end{minipage}
    \caption{Surface brightness decrement as a function of height
      above the plane of the sky for a variety of cavity radii. Each
      curve is labeled with the corresponding \rlos. The curves
      furthest to the left are for the minimum \rlos\ needed to
      reproduce $y_{\rm{min}}$, \ie\ the case of $z = 0$, and the
      horizontal dashed line denotes the minimum decrement for each
      cavity. {\it{Left}} Cavity E1; {\it{Right}} Cavity W1.}
    \label{fig:decs}
  \end{center}
\end{figure}

It is also possible that the cavities are inflating close to our
line-of-sight and thus may be large, elongated structures as shown in
configuration-2 of Figure \ref{fig:config}. Presented in Figure
\ref{fig:decs} are curves showing how surface brightness decrement
changes as a function of $z$ for various \rlos. The plots demonstrate
that it is possible to reproduce the measured E1 and W1 decrements
using larger cavities that have centers displaced from the plane of
the sky. Consequently, the limiting case of $z = 0$ is a lower-limit
on the AGN outburst energetics. If the cavity system lies close to our
line-of-sight, and is much larger and more complex than the data
allows us to constrain, it may explain the additional ICM
substructures seen in the residual images and the ambiguous
relationship between the X-ray and radio morphologies.

\begin{figure}
  \begin{center}
    \begin{minipage}{\linewidth}
      \includegraphics*[width=\textwidth, trim=15mm 5mm 5mm 10mm, clip]{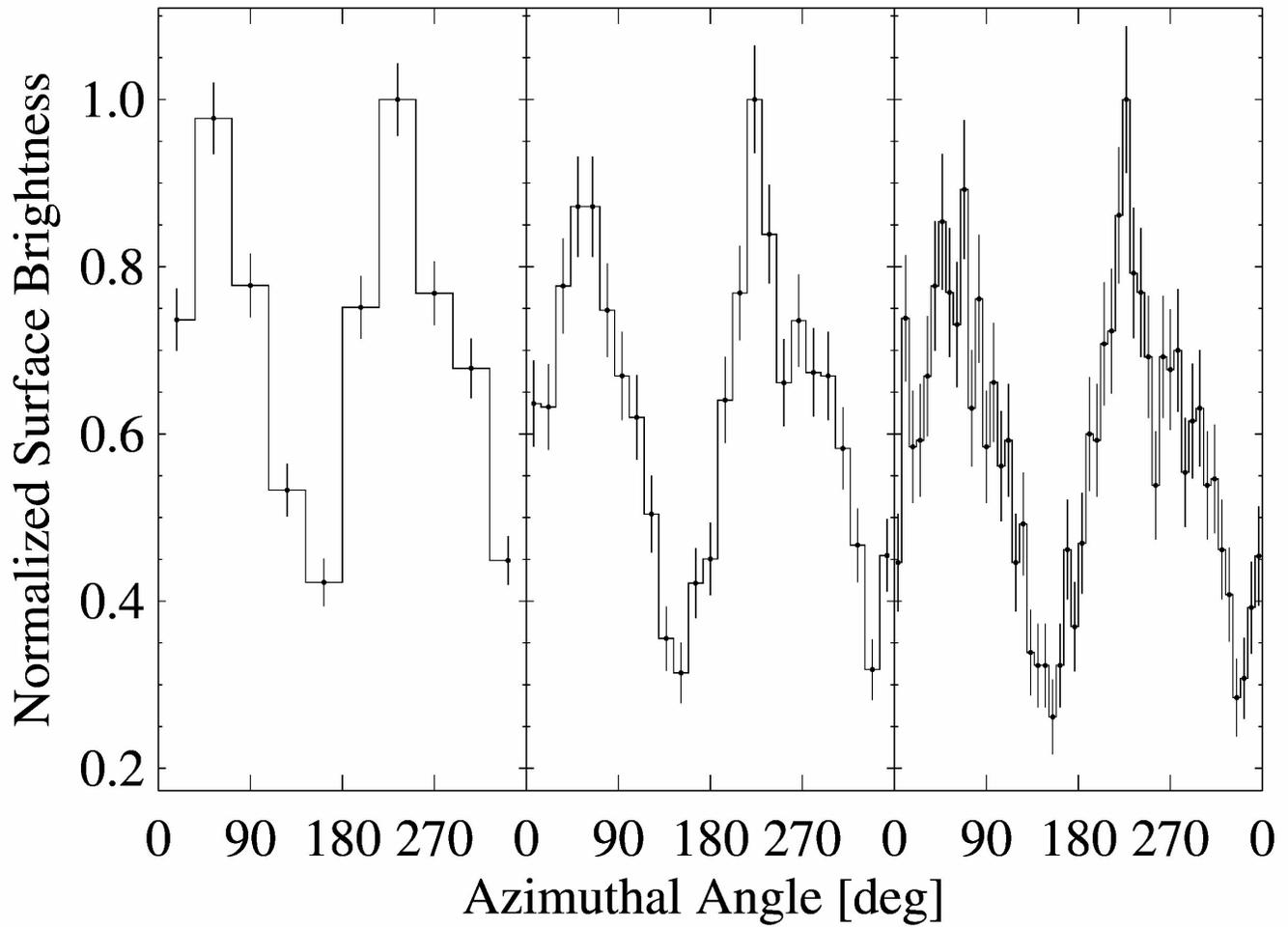}
      \caption{Histograms of normalized surface brightness variation
        in wedges of a $2.5\arcs$ wide annulus centered on the X-ray
        peak and passing through the cavity midpoints. {\it{Left:}}
        $36\mydeg$ wedges; {\it{Middle:}} $14.4\mydeg$ wedges;
        {\it{Right:}} $7.2\mydeg$ wedges. The depth of the cavities
        and prominence of the rims can be clearly seen in this plot.}
      \label{fig:pannorm}
    \end{minipage}
  \end{center}
\end{figure}

It should be noted that deep cavities like in \rbs\ are uncommon, with
most other cavity systems having minima $y \ga 0.6$ (B04). The large
decrements raise the concern that the $\beta$-model fit has been
influenced by the rim-like structures of E1-W1 and produced
artificially large decrements (small $y$). The prominent rims can be
seen in Figure \ref{fig:pannorm} which shows the normalized surface
brightness variation in wedges of a $2.5\arcs$ wide annulus centered
on the X-ray peak and passing through the cavity midpoints. In
addition to excluding the cavities, we tried excluding the rims during
$\beta$-model fitting, but this did not provide insight as too much of
the surface brightness profile was removed and the fit did not
converge. Extrapolating the surface brightness profile at larger radii
inward resulted in even lower decrements. The \citet{hydraa} decrement
model assumes a cavity replaces some X-ray emitting gas with
non-emitting material, without disturbing ambient gas. To be valid,
the model requires that there is little uplifted gas surrounding a
cavity \citep[\cf][]{2005ApJ...628..629N, 2001ApJ...558L..15B} and
that the cavity is not driving significant shocks into its
surroundings (see Section \ref{sec:shocks}). The rims, and their
possible connection to gas shocking, are discussed below.

\subsubsection{Do the Rims Indicate Shocks?}
\label{sec:shocks}

The Chandra X-ray image of \rbs\ bears a strong resemblance to that of
\ms, with a bright elliptical region at its center enclosing two
prominent cavities. In \ms, the bright central region is bounded by
Mach $\simeq 1.4$ shocks. Although there is no evidence for density or
temperature jumps in the Chandra data for \rbs, the spatial resolution
of the \rbs\ observations is poorer than for \ms\ (the angular scale
of shocks in \rbs\ would be $\sim 10\times$ smaller). Based on their
similar appearances, we suggest that the prominent central ellipse in
the X-ray image of \rbs\ may also be surrounded by moderately strong
shocks, and the lack of evidence for shocks may simply be a
consequence of the poor spatial resolution.

A simple model that captures how a bright, shocked rim helps enhance a
cavity decrement is to assume the shocked region is cylindrical and
uniformly compressed. Ignoring emission outside the shocked cylinder,
a compression $\chi$ gives a decrement of $y = \sqrt{\chi} -
\sqrt{\chi - 1}$, hence a decrement of $\ge 0.44$ can be obtained with
$\chi \la 1.84$, \ie\ not dissimilar from the shock Mach numbers in
Hercules A and \ms. This is a rough estimate (arguably pessimistic
since the compressed gas is assumed to be uniform), but it
demonstrates how bright rims created by shocks may enhance the cavity
decrements. Assuming axial symmetry for the unshocked gas, including
emission outside the cylinder adds more to the bright rim than the
cavity center, increasing the surface brightness difference from rim
to cavity, but generally reducing $y$. Because the ICM surface
brightness drops rapidly outside the shocked region, the correction
due to emission from outside the cylinder is modest, and it scales as
$\chi^{-3/2}$, so its effect decreases as the compression
increases. Thus the analogy with \ms\ is also consistent with the
relatively large cavity decrements found in \rbs.

It is often assumed that cavity energetics calculations, like those in
Section \ref{sec:ecav}, provide a reasonable estimate of the physical
quantity jet power, \pjet. However, \pcav\ and \pjet\ do not account
for AGN output energy which may be channeled into shocks. If
significant ICM shocking has occurred in \rbs, then the cavity
energetics clearly underestimate the AGN output, which directly
impacts the discussion of AGN fueling in Section
\ref{sec:accretion}. In the cases where cavity and shock energies have
been directly compared, they are in general comparable. Thus the
impact of shocks on the power estimates may be modest.

\subsection{Powering the Outburst}
\label{sec:accretion}

If the AGN was powered by mass accretion alone, then \ecav\ gives a
lower limit to the gravitational binding energy released as mass
accreted onto the SMBH. Then, the total mass accreted can be
approximated as $\macc = \ecav/(\epsilon c^2)$ with an average
accretion rate $\dmacc = \macc/\tbuoy$. Consequently, the black hole's
mass grew by $\ddmbh = (1-\epsilon)\macc$ at an average rate $\dmbh =
\ddmbh/\tbuoy$. Here, $\epsilon$ is the mass-energy conversion factor
and we adopt the commonly used value $\epsilon = 0.1$
\citep{2002apa..book.....F}. The accretion properties associated with
each cavity configuration are given in Table \ref{tab:totals}, and
\macc\ lies in the range $2 \dash 3 \times 10^7 \msol$. Below, we
consider if the accretion of cold or hot gas pervading the BCG could
meet these requirements without being in conflict with observed BCG
and ICM properties.

\subsubsection{Cold Accretion}
\label{sec:cold}

Optical nebulae and substantial quantities of cold molecular gas are
found in many cool core clusters \citep{crawford99, edge01}. This gas
may be the end point of thermal instabilities in the hot atmosphere,
and potentially are a significant source of fuel for AGN activity and
star formation \citep[\eg][]{pizzolato05, 2006NewA...12...38S,
  2010MNRAS.408..961P}. We are unaware of a molecular gas mass (\mmol)
measurement for \rbs, so we estimate it using the
\mmol-\halpha\ correlation in \citet{edge01}. An optical spectrum
(3200-7600 \AA) of the \rbs\ BCG reveals a strong \hbeta\ emission
line \citep{rbs1, rbs2}, which we used as a surrogate for \halpha\ by
assuming a Balmer decrement of EW$_{\hbeta}$/EW$_{\halpha}$ = 0.29
\citep{2006ApJ...642..775M}, where EW is line equivalent width. We
estimate $\mmol \sim 10^{10} ~\msol$, which is sufficiently in excess
of the $\sim 10^7 ~\msol$ needed to power the outburst. Thus, there is
reason to believe that an ample cold gas reservoir is available to
fuel the AGN. It must be noted that the \mmol-\halpha\ relation has
substantial scatter \citep{salome03}, and that the \rbs\ emission line
measurements are highly uncertain (A. Schwope, private communication),
so the \mmol\ estimate is simply a crude estimate.

Accreting gas from the reservoir will cause the central SMBH mass to
grow, but, on average, black hole mass growth is coupled to the star
formation properties of the host galaxy
\citep[\eg][]{1995ARA&A..33..581K, 2000ApJ...539L...9F}. Assuming star
formation accompanied the phase of accretion which powered the AGN
outburst, the Magorrian relation \citep{magorrian} implies that for
each unit of black hole mass growth, several hundred times as much
goes into bulge stars \citep[\eg][]{2004ApJ...604L..89H}. Thus, the
$\sim 1 ~\msolpy$ mass accretion rate needed to power the
\rbs\ outburst suggests that $\sim 700 ~\msolpy$ of star formation
would be required to grow the galaxy and its SMBH along the Magorrian
relation. The present BCG star formation rate (SFR) is $\sim 1$--10
\msolpy\ (see Section \ref{sec:bcg}), implying that, if the SFR
preceding and during the AGN outburst was of the order the present
rate, the SMBH has grown faster than the slope of the Magorrian
relation implies.

\subsubsection{Hot Accretion}

Direct accretion of the hot ICM via the Bondi mechanism provides
another possible AGN fuel source. The accretion flow arising from this
process is characterized by the Bondi equation (Equation
\ref{eqn:bon}) and is often compared with the Eddington limit
describing the maximal accretion rate for a SMBH (Equation
\ref{eqn:edd}):
\begin{eqnarray}
  \dmbon &=& 0.013 ~\kbon^{-3/2} \left(\frac{\mbh}{10^9
    ~\msol}\right)^{2} ~\msolpy \label{eqn:bon}\\
  \dmedd &=& \frac{2.2}{\epsilon} \left(\frac{\mbh}{10^9~\msol}\right)
  ~\msolpy  \label{eqn:edd}
\end{eqnarray}
where $\epsilon$ is a mass-energy conversion factor, \kbon\ is the
mean entropy [\ent] of gas within the Bondi radius, and \mbh\ is black
hole mass [\msol]. We chose the relations of
\citet{2002ApJ...574..740T} and \citet{2007MNRAS.379..711G} to
estimate \mbh, and find a range of $0.6 \dash 7.8 \times 10^9 ~\msol$,
from which we adopted the weighted mean value $1.5^{+6.3}_{-0.9}
\times 10^9 ~\msol$ where the errors span the lowest and highest
$1\sigma$ values of the individual estimates. For $\epsilon = 0.1$ and
$\kbon = \kna$, the relevant accretion rates are $\dmbon \approx 4
\times 10^{-4} ~\msolpy$ and $\dmedd \approx 33 ~\msolpy$. The
Eddington and Bondi accretion ratios are given in Table
\ref{tab:totals}, with $\dme \equiv \dmacc/\dmedd \approx 0.02$ and
$\dmb \equiv \dmacc/\dmbon \approx 2000 \dash 4000$.

The value of \dmb\ integrated over the duration of the AGN outburst is
apparently larger than the available hot gas supply near the Bondi
radius ($\approx 10$ pc). Using the upper limit of \mbh, $\approx 8
\times 10^9 ~\msol$, and assuming gas near \rbon\ has a mean
temperature of 1.0 keV, a \dmb\ of unity requires \kbon\ be less than
$0.9 ~\ent$, corresponding to $\nelec \approx 1.0 ~\pcc$. This is nine
times the measured ICM central density, although the gas density near
the Bondi radius could be considerably higher and evade detection.
Nevertheless, a sphere more than 1 kpc in radius would required to
house up to $10^7 ~\msol$ needed to power the AGN. These are
properties of a galactic corona \citep{coronae}. A $\sim 1$ kpc, $\tx
\sim 1$ keV, $\nelec \sim 0.4 ~\pcc$ corona would be easily detected
in the \cxo\ observations as a bright point source with a distinct
thermal spectrum. However, the spectrum of the observed nuclear X-ray
point source (see Section \ref{sec:nuc}) is inconsistent with being
thermal in origin (\ie\ no $E < 2$ keV Fe L-shell emission hump and no
$E \approx 6.5$ keV Fe K-shell emission line blend). If the pervading
ICM is instead the source of accretion fuel, these numbers imply the
inner-core of the cluster would have been fully consumed during the
outburst ($<$ 100 Myr), which seems unlikely.

Because the nucleus is unresolved, the gas density at the Bondi radius
could be much higher, particularly if enhanced Compton cooling is
significant \citep[\eg][]{2010MNRAS.402.1561R}. Furthermore, it is
possible that the central gas density was much higher in the past, as
the AGN was turning on. Therefore, Bondi accretion cannot be ruled
out. However, \rbs, like other powerful AGN in clusters, would only be
able to power their AGN by Bondi accretion with great difficulty
\citep{rafferty06, minaspin}.

\subsubsection{Black Hole Spin}

In an effort to find lower accretion rates which still produce
powerful jets, we consider below if a rapidly-spinning SMBH could act
as an alternate power source. We consider the spin model with the
caveats that, like cold- and hot-mode accretion, spin is wrought with
its own difficulties, is hard to evaluate for any one system, and how
rapidly spinning black holes are created is unclear (see
\citealt{msspin} and \citealt{minaspin} for discussion on these
points).

In hybrid spin models \citep[\eg][]{1999ApJ...522..753M,
  2001ApJ...548L...9M, 2006ApJ...651.1023R, 2007MNRAS.377.1652N,
  2009MNRAS.397.1302B, gesspin} jets are produced by a combination of
the Blandford-Znajek \citep{bz} and Blandford-Payne \citep{bp}
mechanisms, \ie\ by extracting energy from a spinning black hole and
its accretion disk via the poloidal component of strong magnetic
fields. In these hybrid models, the accretion rate sets an upper limit
to the strength of the magnetic field that can thread the inner disk
and the black hole, which, in turn, determines the emergent jet
power. Of the model parameters, only \pjet\ is truly measured, so a
range of \dme\ and dimensionless spin values ($j$) can produce any
particular \pjet. However, to avoid the need for excessively large
accretion rates, which is the shortcoming of pure mass accretion
mechanisms, $j$ needs to be as near unity as possible, \ie\ the choice
of \dme\ is not completely arbitrary.

In the hybrid spin model of \citet{2007MNRAS.377.1652N}, the jet
efficiency ($\epsilon$ above) is strongly dependent on black hole
spin. For example, for a disk viscosity parameter of $\alpha = 0.25$,
we require a black hole spin parameter of $j \ga 0.96$ to obtain
$\epsilon > 0.1$. In general, a high jet efficiency demands a high
spin parameter. Alternatively, for $j = 0.7$ the jet efficiency would
be $\approx 0.01$, boosting the mass required to fuel the outburst by
an order of magnitude and exacerbating the issues of accounting for
the fuel discussed above. Additional relief can be found in the model
of \citet{gesspin} which has the feature that extremely powerful jets
are produced when the material in the disk is orbiting retrograde
relative to the spin of the SMBH. In the \citet{gesspin} model, for
$\|j\| \ge 0.9$ and $\mbh \ge 1.5 \times 10^9 ~\msol$, the required
mass accretion rates are $\dme \le 0.005$ or $\la 0.1 ~\msolpy$.

\subsection{Nuclear Emission}
\label{sec:nuc}

\begin{figure}
  \begin{center}
    \begin{minipage}{0.5\linewidth}
      \includegraphics*[width=\textwidth, angle=-90]{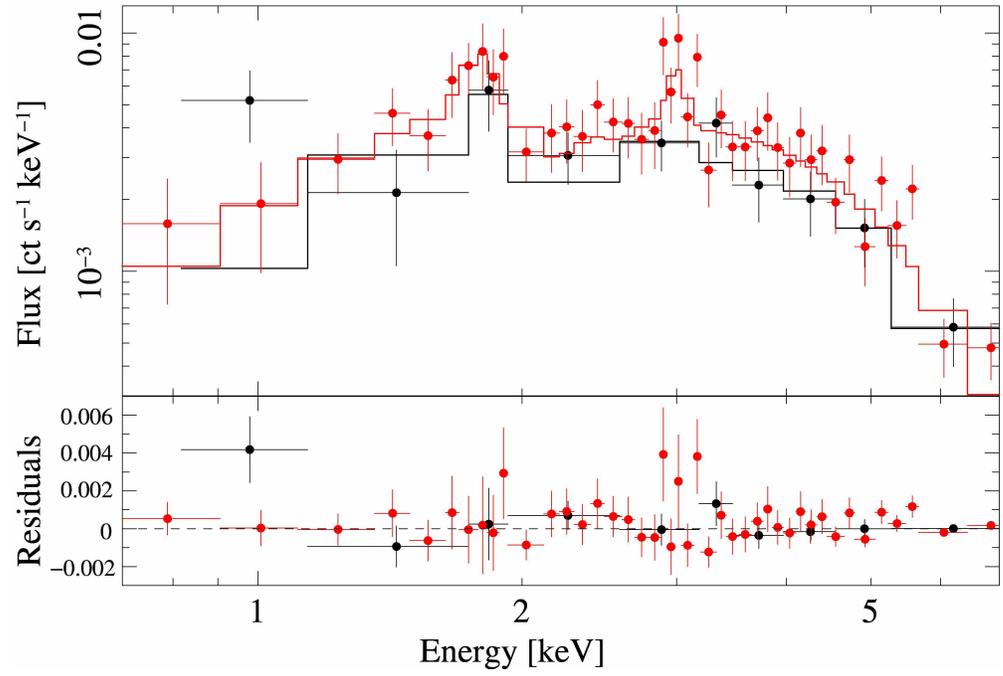}
    \end{minipage}
    \caption{X-ray spectrum of nuclear point source. Black denotes
      year 2000 \cxo\ data (points) and best-fit model (line), and red
      denotes year 2007 \cxo\ data (points) and best-fit model (line).
      The residuals of the fit for both datasets are given below.}
    \label{fig:nucspec}
  \end{center}
\end{figure}

There is a bright BCG X-ray point source apparent in the Chandra image
that coincides with the nucleus of the BCG, and its properties may
also provide clues regarding on-going accretion processes. A source
spectrum was extracted from a region enclosing 90\% of the normalized
\cxo\ PSF specific to the nuclear source median photon energy and
off-axis position. The source region had an effective radius of
$0.86\arcs$, and a background spectrum was taken from an enclosing
annulus that had 5 times the area. The shape and features of the
background-subtracted spectrum, shown in Figure \ref{fig:nucspec}, are
inconsistent with thermal emission. The spectrum was modeled using an
absorbed power law with two Gaussians added to account for emission
features around 1.8 keV and 3.0 keV. Including an absorption edge
component to further correct for the ACIS 2.0 keV iridium feature did
not significantly improve any fit, resulted in optical depths
consistent with zero, and did not negate the need for a Gaussian
around 1.8 keV. For the entire bandpass of both \cxo\ observations,
the number of readout frames exposed to two or more X-ray photons
(\ie\ the pile-up percentage) inside an aperture twice the size of the
source region was $< 6\%$, and addition of a pileup component to the
spectral analysis resulted in no significant improvements to the fits.

A variety of absorption models were fit to the spectrum, and the
best-fit values are given in Table \ref{tab:agn}. The model with a
power-law distribution of $\nhobs \sim 10^{22} ~\pcmsq$ absorbers
yielded the best statistical fit, and the low column densities
indicate the nucleus may not be heavily obscured. If the Gaussian
components represent emission line blends, they would be consistent
with ion species of sulfur, silicon, argon and calcium, possibly
indicating emission or reflection from dense, ionized material in or
near the nucleus \citep[\eg][]{1990ApJ...362...90B,
  1998MNRAS.297.1219I}.

If mass accretion is powering the nuclear emission, the X-ray source
bolometric luminosity, $\lbol = 2.3 \times 10^{44} ~\lum$, implies an
accretion rate of $\dmacc \approx \lbol/(0.1 c^2) \approx 0.04
~\msolpy \approx 0.001 \dme$, well into the regime where accretion
flows are expected to be radiatively inefficient,
\eg\ advection-dominated flows \citep{adaf}. Depending upon the exact
properties of the accretion disk, for $\mbh > 10^9 ~\msol$, the
\hbeta\ luminosity of a $\dme \sim 0.001$ advection-dominated
accretion disk is $\la 10^{42} ~\lum$ \citep{2002ApJ...570L..13C},
which, for a point source, is too faint to be detected against the
extended emission of the BCG nucleus (see Section \ref{sec:bcg}). The
low column densities suggested by the X-ray modeling could not conceal
a $> 10^{42} ~\lum$ source, so if there is a bright, undetected
optical point source, the emission must be beamed away from us.

\begin{figure}
  \begin{center}
    \begin{minipage}{\linewidth}
      \includegraphics*[width=\textwidth, trim=10mm 5mm 10mm 10mm, clip]{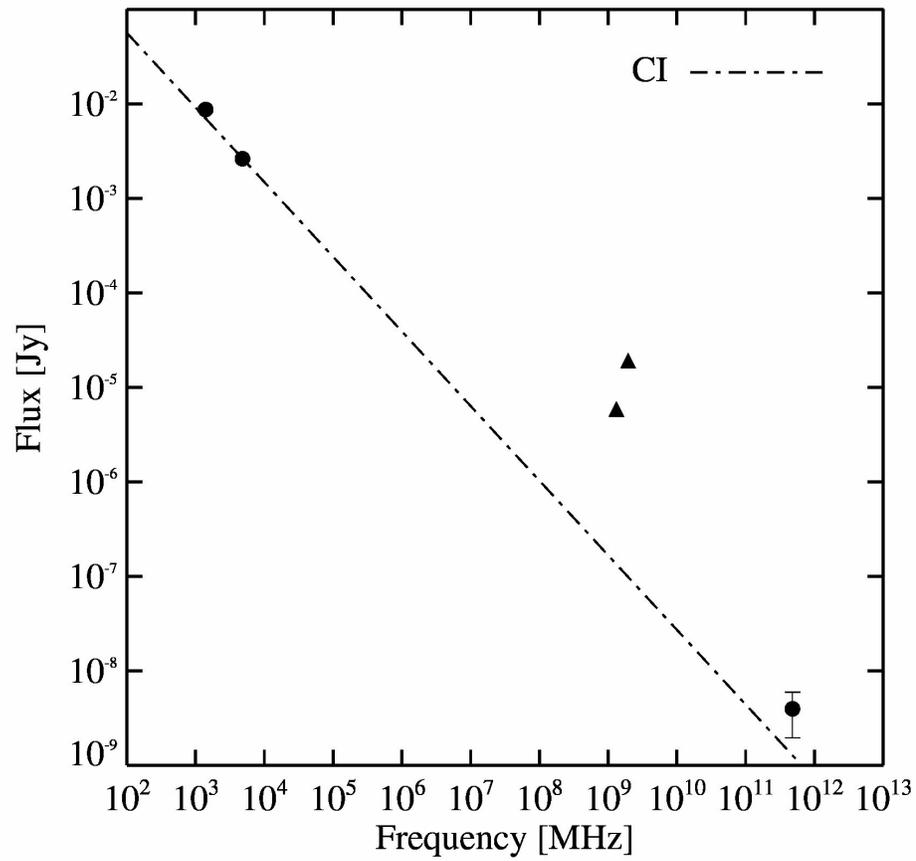}
    \end{minipage}
    \caption{Best-fit continuous injection (CI) synchrotron model to
      the nuclear 1.4 GHz, 4.8 GHz, and 7.0 keV X-ray emission. The
      two triangles are \galex\ UV fluxes showing the emission is
      boosted above the power-law attributable to the nucleus.}
    \label{fig:sync}
    \end{center}
\end{figure}

Extrapolation of the best-fit X-ray spectral model to radio
frequencies reveals good agreement with the measured 1.4 GHz and 4.8
GHz nuclear radio fluxes. Further, the continuous injection
synchrotron model of \citet{1987MNRAS.225..335H} produced an
acceptable fit to the X-ray, 1.4 GHz, and 4.8 GHz nuclear fluxes (see
Figure \ref{fig:sync}). These results suggest the nuclear X-ray source
may be unresolved synchrotron emission from the jets which are obvious
in the high-resolution 4.8 GHz radio image. Regardless, there is no
indication the X-ray source is the remnant of a very dense, hot gas
phase which might be associated with a prior or on-going accretion
event.

\subsection{AGN-BCG Interaction and Constraints on Star Formation}
\label{sec:bcg}

\begin{figure}
  \begin{center}
    \begin{minipage}{\linewidth}
      \includegraphics*[width=\textwidth, trim=0mm 0mm 0mm 0mm, clip]{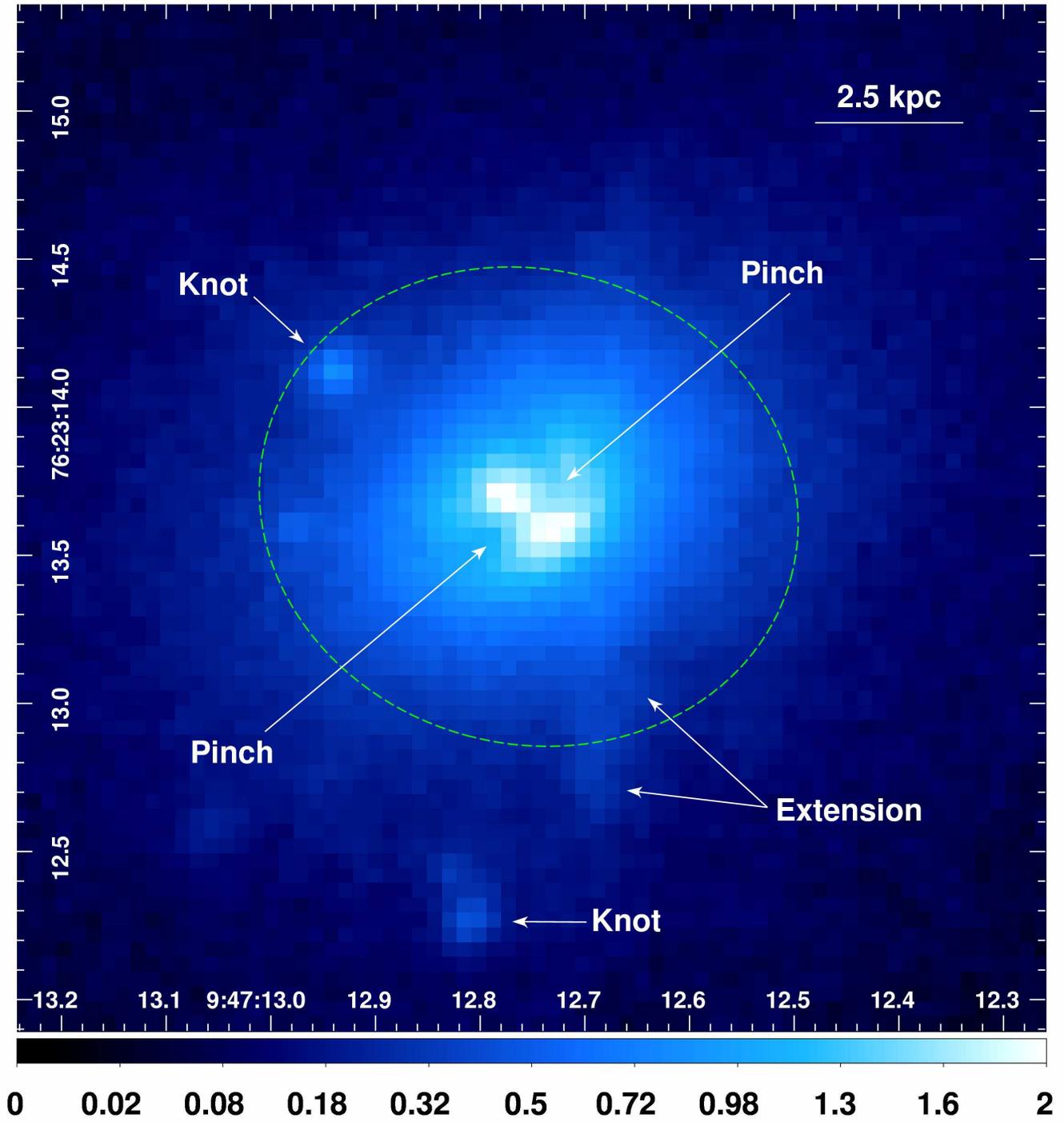}
    \end{minipage}
    \caption{\hst\ \myi+\myv\ image of the \rbs\ BCG with units e$^-$
      s$^{-1}$. Green, dashed contour is the \cxo\ 90\% EEF. Emission
      features discussed in the text are labeled.}
    \label{fig:hst}
  \end{center}
\end{figure}

The BCG's stellar structure and star formation properties were
investigated using archival observations from the Hubble Space
Telescope (\hst), \galex, and the \xmm/Optical Monitor
(\xom). \hst\ imaged \rbs\ using the ACS/WFC instrument and the F606W
(4500--7500 \AA; \myv) and F814W (6800--9800 \AA; \myi)
filters. Images produced using the Hubble Legacy Archive pipeline
version 1.0 were used for analysis. The \hst\ \myv+\myi\ image is
shown in Figure \ref{fig:hst}. Comparison of the \hst\ and
\cxo\ images reveals the BCG coincides with the nuclear X-ray
source. The BCG has the appearance of being bifurcated, perhaps due to
a dust lane, and there is no evidence of an optical point
source. There are also several distinct knots of emission in the halo,
and a faint arm of emission extending south from the BCG. There are
two ACS artifacts in the \myi\ image which begin at $\approx 2.7\arcs$
and $\approx 5.4\arcs$ from the BCG center. Within a $2\arcs$ aperture
centered on the BCG, the measured magnitudes are $m_{\myv} = 19.3 \pm
0.7$ mag and $m_{\myi} = 18.2 \pm 0.6$ mag, consistent with
non-\hst\ measurements of \citet{rbs1}, indicating the
\hst\ photometry in this region is unaffected.

\rbs\ was imaged in the far-UV (FUV; 1344--1786 \AA) and near-UV (NUV;
1771--2831 \AA) with \galex, and the near-UV with the
\xmm/Optical-Monitor (\xom) using filters UVW1 (2410--3565 \AA) and
UVM2 (1970--2675 \AA). The pipeline-reduced observations of
\galex\ Data Release 5 were used for analysis, and the \xom\ data was
processed using \sas\ version 8.0.1. \rbs\ is detected in all but the
UVM2 observation as an unresolved source co-spatial with the optical
and X-ray BCG emission. The individual filter fluxes are $f_{\rm{FUV}}
= 19.2 \pm 4.8 ~\mu$Jy, $f_{\rm{NUV}} = 5.9 \pm 2.1 ~\mu$Jy,
$f_{\rm{UVW1}} = 14.6 \pm 4.6$ $\mu$Jy and $f_{\rm{UVM2}} < 117$
$\mu$Jy, all of which lie above the nuclear power-law emission we
attribute to unresolved jets (see Section \ref{sec:nuc} and Figure
\ref{fig:sync}).

A radial $(\myv-\myi)$ color profile of the central $2\arcs$ ($\approx
10$ kpc) was extracted from the \hst\ images and fitted with the
function $\Delta(\myv-\myi) \log r + b$, where $\Delta(\myv-\myi)$
[mag dex$^{-1}$] is the color gradient, $r$ is radius, and $b$ [mag]
is a normalization. The best-fit parameters are $\Delta(\myv-\myi) =
-0.20 \pm 0.02$ and $b = 1.1 \pm 0.01$ for \chisq(DOF) = 0.009(21),
revealing a flat color profile.

BCGs, and elliptical galaxies in general, have red centers and bluer
halos due to the higher metallicity of the central stellar populations
\citep{1990ApJS...73..637M}. BCGs in cooling flows with significant
star formation have unusually blue cores relative to the quiescent
halos \citep[\eg][]{rafferty06}. Our analysis shows a flat color
profile, which is consistent with a modest level of nuclear star
formation, but is inconsistent with star formation occurring at
several tens of solar masses per year. Lacking calibrated colors, the
color gradient analysis alone is ambiguous, since other factors, such
as dust, emission line contamination in the passbands, and metallicity
variations can alter the profile slope.

Star formation rates were estimated from the UV fluxes using the
relations of \citet{kennicutt2} and \citet{salim2007}. We estimate
rates in the range 1--10 ~\msolpy\ (see Table \ref{tab:sfr}), which
are probably consistent with the flat color profile. These estimates
should be considered upper limits because sources of significant
uncertainty, such as AGN contamination, have been neglected.

\begin{figure}
  \begin{center}
    \begin{minipage}{0.495\linewidth}
      \includegraphics*[width=\textwidth, trim=0mm 0mm 0mm 0mm, clip]{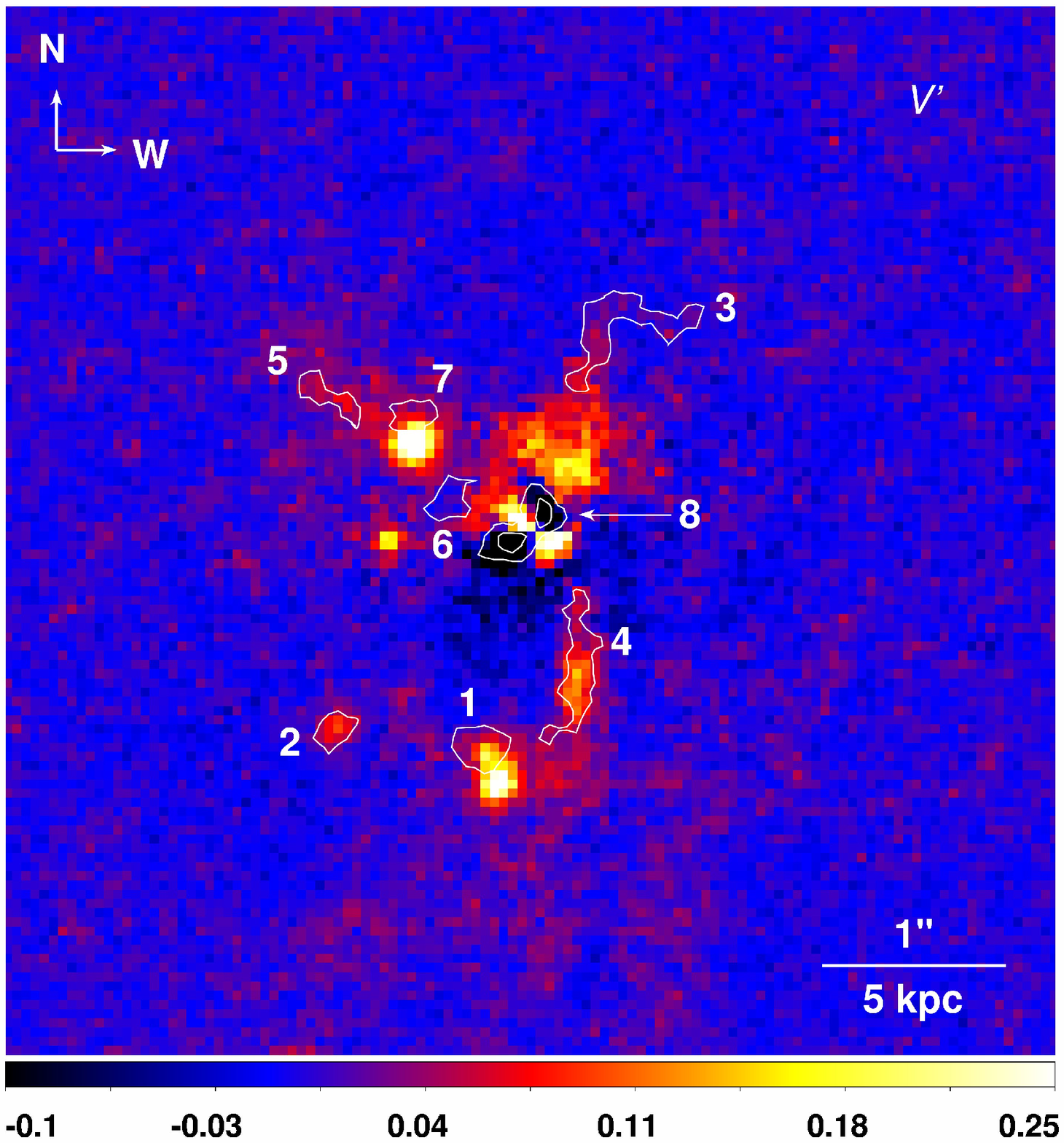}
    \end{minipage}
    \begin{minipage}{0.495\linewidth}
      \includegraphics*[width=\textwidth, trim=0mm 0mm 0mm 0mm, clip]{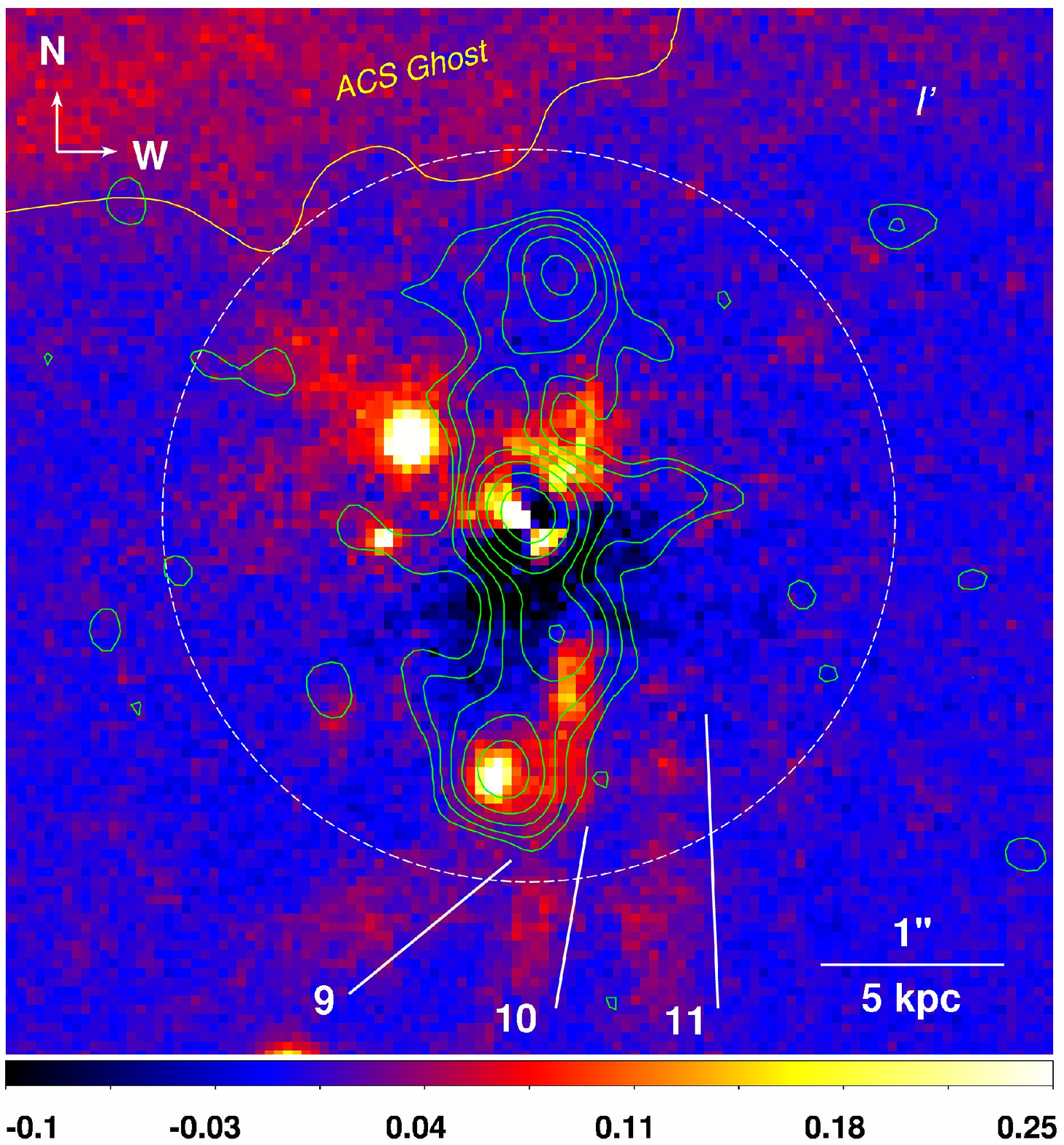}
    \end{minipage}
    \caption{{\it{Left:}} Residual \hst\ \myv\ image. White regions
      (numbered 1--8) are areas with greatest color difference with
      \rbs\ halo. {\it{Right:}} Residual \hst\ \myi\ image. Green
      contours are 4.8 GHz radio emission down to
      $1\sigma_{\rm{rms}}$, white dashed circle has radius $2\arcs$,
      edge of ACS ghost is show in yellow, and southern whiskers are
      numbered 9--11 with corresponding white lines.}
    \label{fig:subopt}
  \end{center}
\end{figure}

The HST images have revealed a great deal of structure in the BCG. In
order to investigate this structure, residual galaxy images were
constructed by first fitting the \hst\ \myv\ and \myi\ isophotes with
ellipses using the \iraf\ tool {\textsc{ellipse}}. Foreground stars
and other contaminating sources were rejected using a combination of
$3\sigma$ clipping and by masking. The ellipse centers were fixed at
the galaxy centroid, and ellipticity and position angle were fixed at
$0.25 \pm 0.02$ and $-64\mydeg \pm 2\mydeg$, respectively -- the mean
values when they were free parameters. Galaxy light models were
created using {\textsc{bmodel}} in \iraf\ and subtracted from the
corresponding parent image, leaving the residual images shown in
Figure \ref{fig:subopt}. A color map was also generated by subtracting
the fluxed \myi\ image from the fluxed \myv\ image.

The close alignment of the optical substructure with the nuclear AGN
outflow clearly indicates the jets are interacting with material in
the BCG's halo. An $\approx 1$ kpc radius spheroid $\approx 4$ kpc
northeast of the nucleus resembles a galaxy which may be falling
through the core and being stripped \citep[see][for
  example]{2007ApJ...671..190S}. The numbered regions overlaid on the
residual \myv\ image are the areas of the color map which have the
largest color difference with surrounding galaxy light. Regions 1--5
are relatively the bluest with $\Delta(\myv-\myi) = -0.40, -0.30,
-0.25, -0.22, ~\rm{and} -0.20$, respectively. Regions 6--8 are
relatively the reddest with $\Delta(\myv-\myi) =$ +0.10, +0.15, and
+0.18, respectively. Clusters like \rbs\ with a core entropy less than
$30 ~\ent$ often host a BCG surrounded by extended \halpha\ nebulae
\citep[\eg][]{mcdonald10}. Like these systems, the residual
\myi\ image reveals what appear to be 8--10 kpc long ``whiskers''
surrounding the BCG (regions 9--11). It is interesting that blue
regions 1 and 4 reside at the point where the southern jet appears to
be encountering whiskers 9 and 10.

It is unclear whether the structure is associated with dusty star
formation, emission line nebulae or both. The influence of optical
emission lines on the photometry was estimated crudely in the nucleus
using the ratio of line EWs taken from \citet{rbs1} to \hst\ filter
widths. The nuclear \halpha\ contribution to the \myi\ image is
estimated at $\approx 3\%$ using the scaled \hbeta\ line (see Section
\ref{sec:cold}), and the combined \hbeta, \oii, and
\oiii\ contribution to the nuclear \myv\ image is $\approx 7\%$. The
observed structures have magnitude differences that significantly
exceed these values, which suggests that emission lines alone may not
be responsible for the structure. However, because the stellar
continuum drops rapidly into the halo of the galaxy, bright knots of
nebular emission would have larger equivalent widths at larger radii,
and thus may be able to explain the extended structure.

\section{Conclusions}
\label{sec:con}

\begin{figure}
  \begin{center}
    \begin{minipage}{0.495\linewidth}
      \includegraphics*[width=\textwidth, trim=0mm 0mm 0mm 0mm, clip]{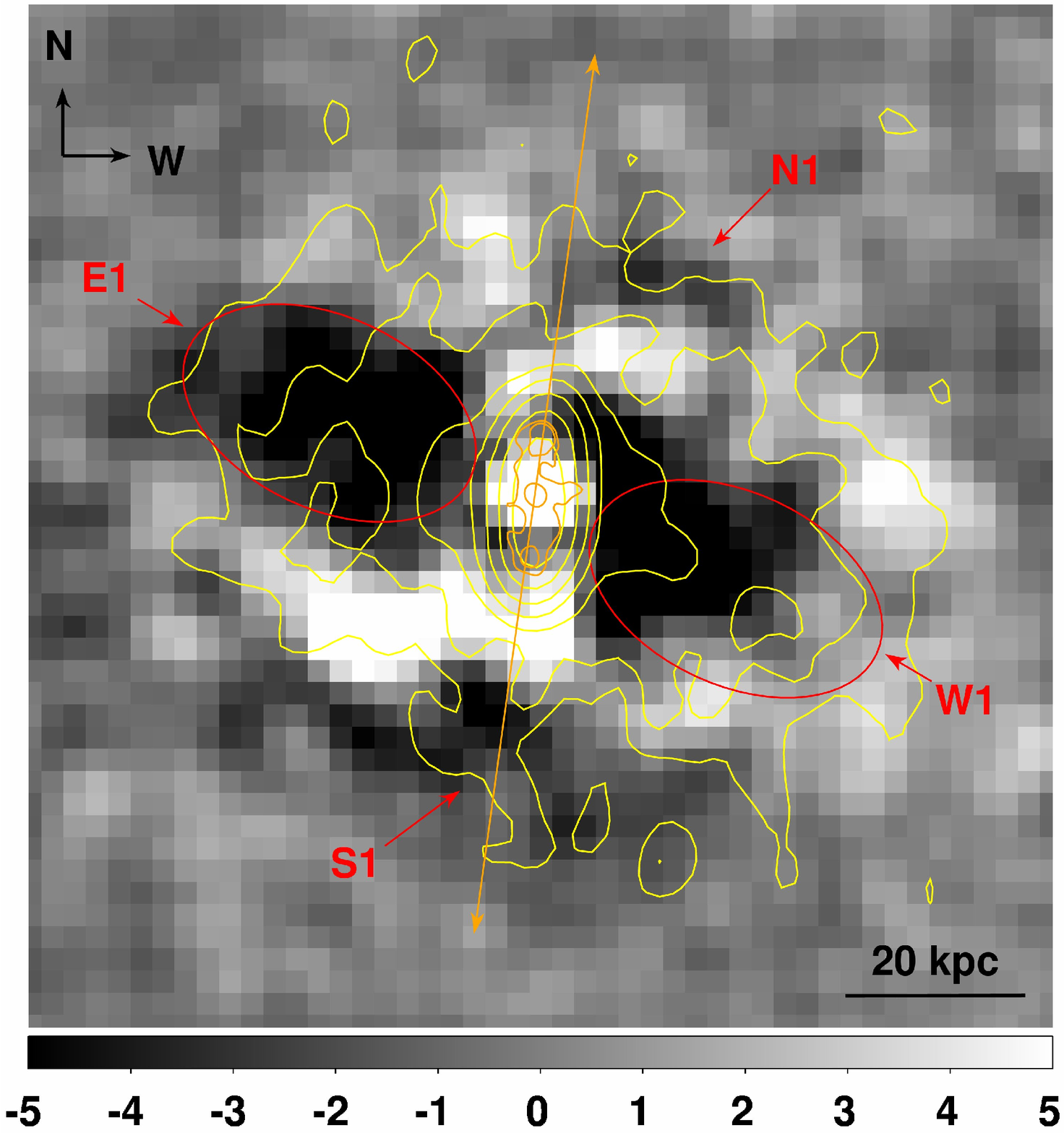}
    \end{minipage}
    \begin{minipage}{0.495\linewidth}
      \includegraphics*[width=\textwidth, trim=0mm 0mm 0mm 0mm, clip]{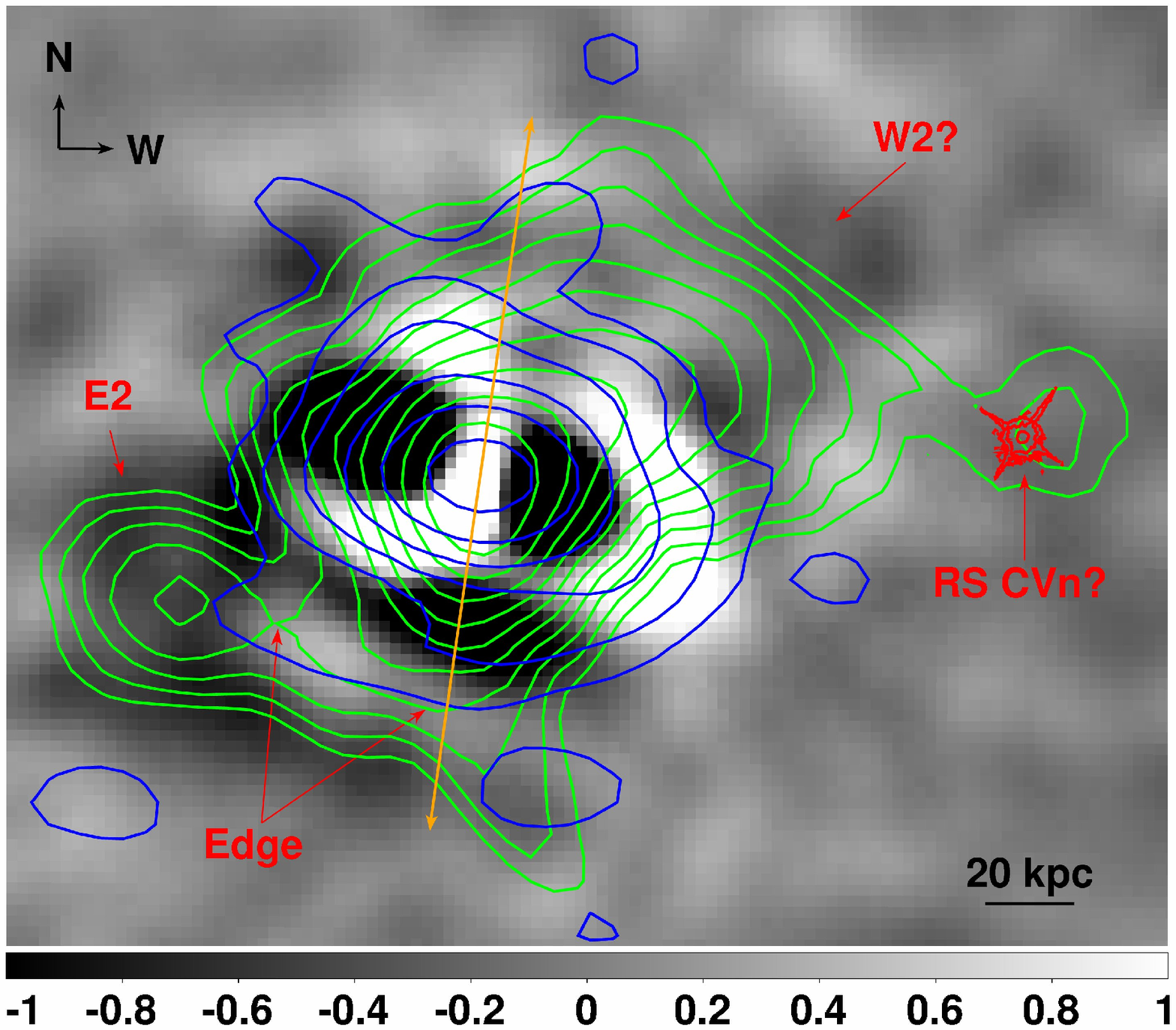}
    \end{minipage}
    \caption{Red text point-out regions of interest discussed in
      Section \ref{sec:cavities}. {\it{Left:}} Residual 0.3-10.0 keV
      X-ray image smoothed with $1\arcs$ Gaussian. Yellow contours are
      1.4 GHz emission (\vla\ A-array), orange contours are 4.8 GHz
      emission (\vla\ A-array), orange vector is 4.8 GHz jet axis, and
      red ellipses outline definite cavities. {\it{Right:}} Residual
      0.3-10.0 keV X-ray image smoothed with $3\arcs$ Gaussian. Green
      contours are 325 MHz emission (\vla\ A-array), blue contours are
      8.4 GHz emission (\vla\ D-array), and orange vector is 4.8 GHz
      jet axis.}
    \label{fig:subxray}
  \end{center}
\end{figure}

We have presented results from a study of the AGN outburst in the
galaxy cluster \rbs. \cxo\ observations have enabled us to constrain
the energetics of the AGN outburst and analyze different powering
mechanisms. We have shown the following:
\begin{enumerate}
\item In addition to the two previously-known cavities near the
  cluster core, residual imaging reveals extensive structure in the
  ICM (Figure \ref{fig:subxray}) associated with the AGN. The ICM
  substructure and deep cavity decrements lead us to speculate that
  the cavity system may be much larger and more complex than the
  present data allows us to constrain.
\item We find that the two central cavities have surface brightness
  decrements that are unusually deep and are inconsistent with the
  cavities being spheres whose centers lie in a plane perpendicular to
  our line of sight which passes through the central AGN. Motivated by
  the decrement analysis, we propose that the cavities are either
  highly-elongated structures in the plane of the sky, or result from
  the superposition of much larger structures lying along our line of
  sight.
\item Using the cavity decrements as a constraint, we estimate the
  total AGN outburst energy at $3 \dash 6 \times 10^{60}$ erg with a
  total power output of $3 \dash 6 \times 10^{45} ~\lum$. The cavities
  may be larger than we have assumed and we consider the energetics
  estimates to be lower limits. The thick, bright rims surrounding the
  cavities may also be signaling the presence of shocks, which would
  further boost the AGN energy output.
\item We show that the AGN can be plausibly powered by accretion of
  cold gas, but accretion of hot gas via the Bondi mechanism is almost
  certainly implausible. We also demonstrate that the outburst could
  be powered by tapping the energy stored in a maximally spinning
  SMBH.
\item Archival \hst\ imagery has revealed a great deal of structure in
  the BCG associated with either star formation, nebular emission, or
  both. The association of the optical structure with the radio lobes
  and X-ray cavities indicates that the AGN is interacting with cooler
  gas in the host galaxy. While we are unable to determine if any
  regions of interest host star formation or line emission, the
  convergence of what appear to be extended optical filaments, bluish
  knots of emission, and the tip of one jet suggest there may be AGN
  driven star formation.
\end{enumerate}

\acknowledgements

KWC acknowledges financial support from L'Agence Nationale de la
Recherche through grant ANR-09-JCJC-0001-01, and thanks David Gilbank,
Sabine Schindler, Axel Schwope, and Chris Waters for helpful
input. BRM was supported by a generous grant from the Canadian Natural
Science and Engineering Research Council, and thanks support provided
by the National Aeronautics and Space Administration through Chandra
Award Number G07-8122X issued by the Chandra X-ray Observatory Center,
which is operated by the Smithsonian Astrophysical Observatory for and
on behalf of the National Aeronautics Space Administration under
contract NAS8-03060. MG acknowledges support from the grants ASI-INAF
I/088/06/0 and CXO GO0-11003X. Some results are based on observations
made with the NASA/ESA Hubble Space Telescope, and obtained from the
Hubble Legacy Archive, which is a collaboration between the Space
Telescope Science Institute (STScI/NASA), the Space Telescope European
Coordinating Facility (ST-ECF/ESA) and the Canadian Astronomy Data
Centre (CADC/NRC/CSA). This research has made use of NASA's
Astrophysics Data System (ADS), Extragalactic Database (NED), High
Energy Astrophysics Science Archive Research Center (HEASARC), data
obtained from the Chandra Data Archive, and software provided by the
Chandra X-ray Center (CXC).

{\it Facilities:} \facility{BAT ()}, \facility{CXO (ACIS)},
\facility{GALEX ()}, \facility{HST (ACS/WFC)}, \facility{VLA ()},
\facility{XMM (OM)}

\bibliography{cavagnolo}

\begin{thebibliography}{71}
\expandafter\ifx\csname natexlab\endcsname\relax\def\natexlab#1{#1}\fi

\bibitem[{{Allen} {et~al.}(2006){Allen}, {Dunn}, {Fabian}, {Taylor}, \&
  {Reynolds}}]{2006MNRAS.372...21A}
{Allen}, S.~W., {Dunn}, R.~J.~H., {Fabian}, A.~C., {Taylor}, G.~B., \&
  {Reynolds}, C.~S. 2006, \mnras, 372, 21

\bibitem[{{Anders} \& {Grevesse}(1989)}]{ag89}
{Anders}, E., \& {Grevesse}, N. 1989, \gca, 53, 197

\bibitem[{{Arnaud}(1996)}]{xspec}
{Arnaud}, K.~A. 1996, in Astronomical Society of the Pacific Conference Series,
  Vol. 101, Astronomical Data Analysis Software and Systems V, ed. G.~H.
  {Jacoby} \& J.~{Barnes}, 17--+

\bibitem[{{Band} {et~al.}(1990){Band}, {Klein}, {Castor}, \&
  {Nash}}]{1990ApJ...362...90B}
{Band}, D.~L., {Klein}, R.~I., {Castor}, J.~I., \& {Nash}, J.~K. 1990, \apj,
  362, 90

\bibitem[{{Benson} \& {Babul}(2009)}]{2009MNRAS.397.1302B}
{Benson}, A.~J., \& {Babul}, A. 2009, \mnras, 397, 1302

\bibitem[{{B{\^\i}rzan} {et~al.}(2008){B{\^\i}rzan}, {McNamara}, {Nulsen},
  {Carilli}, \& {Wise}}]{birzan08}
{B{\^\i}rzan}, L., {McNamara}, B.~R., {Nulsen}, P.~E.~J., {Carilli}, C.~L., \&
  {Wise}, M.~W. 2008, \apj, 686, 859

\bibitem[{{B{\^\i}rzan} {et~al.}(2004){B{\^\i}rzan}, {Rafferty}, {McNamara},
  {Wise}, \& {Nulsen}}]{birzan04}
{B{\^\i}rzan}, L., {Rafferty}, D.~A., {McNamara}, B.~R., {Wise}, M.~W., \&
  {Nulsen}, P.~E.~J. 2004, \apj, 607, 800

\bibitem[{{Blandford} \& {Payne}(1982)}]{bp}
{Blandford}, R.~D., \& {Payne}, D.~G. 1982, \mnras, 199, 883

\bibitem[{{Blandford} \& {Znajek}(1977)}]{bz}
{Blandford}, R.~D., \& {Znajek}, R.~L. 1977, \mnras, 179, 433

\bibitem[{{Blanton} {et~al.}(2001){Blanton}, {Sarazin}, {McNamara}, \&
  {Wise}}]{2001ApJ...558L..15B}
{Blanton}, E.~L., {Sarazin}, C.~L., {McNamara}, B.~R., \& {Wise}, M.~W. 2001,
  \apjl, 558, L15

\bibitem[{{Bower} {et~al.}(2006){Bower}, {Benson}, {Malbon}, {Helly}, {Frenk},
  {Baugh}, {Cole}, \& {Lacey}}]{bower06}
{Bower}, R.~G., {Benson}, A.~J., {Malbon}, R., {Helly}, J.~C., {Frenk}, C.~S.,
  {Baugh}, C.~M., {Cole}, S., \& {Lacey}, C.~G. 2006, \mnras, 370, 645

\bibitem[{{Cao}(2002)}]{2002ApJ...570L..13C}
{Cao}, X. 2002, \apjl, 570, L13

\bibitem[{{Cassano} {et~al.}(2008){Cassano}, {Gitti}, \&
  {Brunetti}}]{2008A&A...486L..31C}
{Cassano}, R., {Gitti}, M., \& {Brunetti}, G. 2008, \aap, 486, L31

\bibitem[{{Cavagnolo} {et~al.}(2008{\natexlab{a}}){Cavagnolo}, {Donahue},
  {Voit}, \& {Sun}}]{haradent}
{Cavagnolo}, K.~W., {Donahue}, M., {Voit}, G.~M., \& {Sun}, M.
  2008{\natexlab{a}}, \apjl, 683, L107

\bibitem[{{Cavagnolo} {et~al.}(2008{\natexlab{b}}){Cavagnolo}, {Donahue},
  {Voit}, \& {Sun}}]{xrayband}
---. 2008{\natexlab{b}}, \apj, 682, 821

\bibitem[{{Cavagnolo} {et~al.}(2009){Cavagnolo}, {Donahue}, {Voit}, \&
  {Sun}}]{accept}
---. 2009, \apjs, 182, 12

\bibitem[{{Crawford} {et~al.}(1999){Crawford}, {Allen}, {Ebeling}, {Edge}, \&
  {Fabian}}]{crawford99}
{Crawford}, C.~S., {Allen}, S.~W., {Ebeling}, H., {Edge}, A.~C., \& {Fabian},
  A.~C. 1999, \mnras, 306, 857

\bibitem[{{Croston} {et~al.}(2010){Croston}, {Hardcastle}, {Mingo}, {Evans},
  {Dicken}, {Morganti}, \& {Tadhunter}}]{2010arXiv1011.6405C}
{Croston}, J.~H., {Hardcastle}, M.~J., {Mingo}, B., {Evans}, D.~A., {Dicken},
  D., {Morganti}, R., \& {Tadhunter}, C.~N. 2010, ArXiv e-prints

\bibitem[{{Croton} {et~al.}(2006){Croton}, {Springel}, {White}, {De Lucia},
  {Frenk}, {Gao}, {Jenkins}, {Kauffmann}, {Navarro}, \& {Yoshida}}]{croton06}
{Croton}, D.~J., {et~al.} 2006, \mnras, 365, 11

\bibitem[{{De Filippis} {et~al.}(2002){De Filippis}, {Schindler}, \&
  {Castillo-Morales}}]{2002astro.ph..1349D}
{De Filippis}, E., {Schindler}, S., \& {Castillo-Morales}, A. 2002, arXiv
  e-prints: 0201349

\bibitem[{{Dunn} \& {Fabian}(2006)}]{dunn06}
{Dunn}, R.~J.~H., \& {Fabian}, A.~C. 2006, \mnras, 373, 959

\bibitem[{{Edge}(2001)}]{edge01}
{Edge}, A.~C. 2001, \mnras, 328, 762

\bibitem[{{Ferrarese} \& {Merritt}(2000)}]{2000ApJ...539L...9F}
{Ferrarese}, L., \& {Merritt}, D. 2000, \apjl, 539, L9

\bibitem[{{Fischer} {et~al.}(1998){Fischer}, {Hasinger}, {Schwope}, {Brunner},
  {Boller}, {Tr{\"u}mper}, {Voges}, \& {Neizvestnyj}}]{rbs1}
{Fischer}, J.-U., {Hasinger}, G., {Schwope}, A.~D., {Brunner}, H., {Boller},
  T., {Tr{\"u}mper}, J., {Voges}, W., \& {Neizvestnyj}, S. 1998, Astronomische
  Nachrichten, 319, 347

\bibitem[{{Frank} {et~al.}(2002){Frank}, {King}, \&
  {Raine}}]{2002apa..book.....F}
{Frank}, J., {King}, A., \& {Raine}, D.~J. 2002, {Accretion Power in
  Astrophysics: Third Edition}, ed. {Frank, J., King, A., \& Raine, D.~J.}

\bibitem[{{Garofalo} {et~al.}(2010){Garofalo}, {Evans}, \&
  {Sambruna}}]{gesspin}
{Garofalo}, D., {Evans}, D.~A., \& {Sambruna}, R.~M. 2010, \mnras, 820

\bibitem[{{Gitti} {et~al.}(2006){Gitti}, {Feretti}, \& {Schindler}}]{gitti06}
{Gitti}, M., {Feretti}, L., \& {Schindler}, S. 2006, \aap, 448, 853

\bibitem[{{Graham}(2007)}]{2007MNRAS.379..711G}
{Graham}, A.~W. 2007, \mnras, 379, 711

\bibitem[{{H{\"a}ring} \& {Rix}(2004)}]{2004ApJ...604L..89H}
{H{\"a}ring}, N., \& {Rix}, H. 2004, \apjl, 604, L89

\bibitem[{{Heavens} \& {Meisenheimer}(1987)}]{1987MNRAS.225..335H}
{Heavens}, A.~F., \& {Meisenheimer}, K. 1987, \mnras, 225, 335

\bibitem[{{Iwasawa} \& {Comastri}(1998)}]{1998MNRAS.297.1219I}
{Iwasawa}, K., \& {Comastri}, A. 1998, \mnras, 297, 1219

\bibitem[{{Kalberla} {et~al.}(2005){Kalberla}, {Burton}, {Hartmann}, {Arnal},
  {Bajaja}, {Morras}, \& {P{\"o}ppel}}]{lab}
{Kalberla}, P.~M.~W., {Burton}, W.~B., {Hartmann}, D., {Arnal}, E.~M.,
  {Bajaja}, E., {Morras}, R., \& {P{\"o}ppel}, W.~G.~L. 2005, \aap, 440, 775

\bibitem[{{Kauffmann} \& {Haehnelt}(2000)}]{2000MNRAS.311..576K}
{Kauffmann}, G., \& {Haehnelt}, M. 2000, \mnras, 311, 576

\bibitem[{{Kennicutt}(1998)}]{kennicutt2}
{Kennicutt}, Jr., R.~C. 1998, \araa, 36, 189

\bibitem[{{Kormendy} \& {Richstone}(1995)}]{1995ARA&A..33..581K}
{Kormendy}, J., \& {Richstone}, D. 1995, \araa, 33, 581

\bibitem[{{Kriss} {et~al.}(1983){Kriss}, {Cioffi}, \& {Canizares}}]{kriss83}
{Kriss}, G.~A., {Cioffi}, D.~F., \& {Canizares}, C.~R. 1983, \apj, 272, 439

\bibitem[{{Mackie} {et~al.}(1990){Mackie}, {Visvanathan}, \&
  {Carter}}]{1990ApJS...73..637M}
{Mackie}, G., {Visvanathan}, N., \& {Carter}, D. 1990, \apjs, 73, 637

\bibitem[{{Magorrian} {et~al.}(1998){Magorrian}, {Tremaine}, {Richstone},
  {Bender}, {Bower}, {Dressler}, {Faber}, {Gebhardt}, {Green}, {Grillmair},
  {Kormendy}, \& {Lauer}}]{magorrian}
{Magorrian}, J., {et~al.} 1998, \aj, 115, 2285

\bibitem[{{McDonald} {et~al.}(2010){McDonald}, {Veilleux}, {Rupke}, \&
  {Mushotzky}}]{mcdonald10}
{McDonald}, M., {Veilleux}, S., {Rupke}, D.~S.~N., \& {Mushotzky}, R. 2010,
  \apj, 721, 1262

\bibitem[{{McNamara} {et~al.}(2009){McNamara}, {Kazemzadeh}, {Rafferty},
  {B{\^\i}rzan}, {Nulsen}, {Kirkpatrick}, \& {Wise}}]{msspin}
{McNamara}, B.~R., {Kazemzadeh}, F., {Rafferty}, D.~A., {B{\^\i}rzan}, L.,
  {Nulsen}, P.~E.~J., {Kirkpatrick}, C.~C., \& {Wise}, M.~W. 2009, \apj, 698,
  594

\bibitem[{{McNamara} \& {Nulsen}(2007)}]{mcnamrev}
{McNamara}, B.~R., \& {Nulsen}, P.~E.~J. 2007, \araa, 45, 117

\bibitem[{{McNamara} {et~al.}(2005){McNamara}, {Nulsen}, {Wise}, {Rafferty},
  {Carilli}, {Sarazin}, \& {Blanton}}]{ms0735}
{McNamara}, B.~R., {Nulsen}, P.~E.~J., {Wise}, M.~W., {Rafferty}, D.~A.,
  {Carilli}, C., {Sarazin}, C.~L., \& {Blanton}, E.~L. 2005, \nat, 433, 45

\bibitem[{{McNamara} {et~al.}(2010){McNamara}, {Rohanizadegan}, \&
  {Nulsen}}]{minaspin}
{McNamara}, B.~R., {Rohanizadegan}, M., \& {Nulsen}, P.~E.~J. 2010, arXiv
  e-prints: 1007.1227

\bibitem[{{Meier}(1999)}]{1999ApJ...522..753M}
{Meier}, D.~L. 1999, \apj, 522, 753

\bibitem[{{Meier}(2001)}]{2001ApJ...548L...9M}
---. 2001, \apjl, 548, L9

\bibitem[{{Mewe} {et~al.}(1985){Mewe}, {Gronenschild}, \& {van den
  Oord}}]{mekal1}
{Mewe}, R., {Gronenschild}, E.~H.~B.~M., \& {van den Oord}, G.~H.~J. 1985,
  \aaps, 62, 197

\bibitem[{{Moustakas} {et~al.}(2006){Moustakas}, {Kennicutt}, \&
  {Tremonti}}]{2006ApJ...642..775M}
{Moustakas}, J., {Kennicutt}, Jr., R.~C., \& {Tremonti}, C.~A. 2006, \apj, 642,
  775

\bibitem[{{Narayan} \& {Yi}(1995)}]{adaf}
{Narayan}, R., \& {Yi}, I. 1995, \apj, 452, 710

\bibitem[{{Nemmen} {et~al.}(2007){Nemmen}, {Bower}, {Babul}, \&
  {Storchi-Bergmann}}]{2007MNRAS.377.1652N}
{Nemmen}, R.~S., {Bower}, R.~G., {Babul}, A., \& {Storchi-Bergmann}, T. 2007,
  \mnras, 377, 1652

\bibitem[{{Nulsen} {et~al.}(2005{\natexlab{a}}){Nulsen}, {Hambrick},
  {McNamara}, {Rafferty}, {B\^irzan}, {Wise}, \& {David}}]{herca}
{Nulsen}, P.~E.~J., {Hambrick}, D.~C., {McNamara}, B.~R., {Rafferty}, D.,
  {B\^irzan}, L., {Wise}, M.~W., \& {David}, L.~P. 2005{\natexlab{a}}, \apjl,
  625, L9

\bibitem[{{Nulsen} {et~al.}(2005{\natexlab{b}}){Nulsen}, {McNamara}, {Wise}, \&
  {David}}]{2005ApJ...628..629N}
{Nulsen}, P.~E.~J., {McNamara}, B.~R., {Wise}, M.~W., \& {David}, L.~P.
  2005{\natexlab{b}}, \apj, 628, 629

\bibitem[{{Peterson} \& {Fabian}(2006)}]{cfreview}
{Peterson}, J.~R., \& {Fabian}, A.~C. 2006, \physrep, 427, 1

\bibitem[{{Pizzolato} \& {Soker}(2005)}]{pizzolato05}
{Pizzolato}, F., \& {Soker}, N. 2005, \apj, 632, 821

\bibitem[{{Pizzolato} \& {Soker}(2010)}]{2010MNRAS.408..961P}
---. 2010, \mnras, 408, 961

\bibitem[{{Rafferty} {et~al.}(2008){Rafferty}, {McNamara}, \&
  {Nulsen}}]{rafferty08}
{Rafferty}, D.~A., {McNamara}, B.~R., \& {Nulsen}, P.~E.~J. 2008, \apj, 687,
  899

\bibitem[{{Rafferty} {et~al.}(2006){Rafferty}, {McNamara}, {Nulsen}, \&
  {Wise}}]{rafferty06}
{Rafferty}, D.~A., {McNamara}, B.~R., {Nulsen}, P.~E.~J., \& {Wise}, M.~W.
  2006, \apj, 652, 216

\bibitem[{{Reynolds} {et~al.}(2006){Reynolds}, {Garofalo}, \&
  {Begelman}}]{2006ApJ...651.1023R}
{Reynolds}, C.~S., {Garofalo}, D., \& {Begelman}, M.~C. 2006, \apj, 651, 1023

\bibitem[{{Russell} {et~al.}(2010){Russell}, {Fabian}, {Sanders}, {Johnstone},
  {Blundell}, {Brandt}, \& {Crawford}}]{2010MNRAS.402.1561R}
{Russell}, H.~R., {Fabian}, A.~C., {Sanders}, J.~S., {Johnstone}, R.~M.,
  {Blundell}, K.~M., {Brandt}, W.~N., \& {Crawford}, C.~S. 2010, \mnras, 402,
  1561

\bibitem[{{Salim} {et~al.}(2007){Salim}, {Rich}, {Charlot}, {Brinchmann},
  {Johnson}, {Schiminovich}, {Seibert}, {Mallery}, {Heckman}, {Forster},
  {Friedman}, {Martin}, {Morrissey}, {Neff}, {Small}, {Wyder}, {Bianchi},
  {Donas}, {Lee}, {Madore}, {Milliard}, {Szalay}, {Welsh}, \& {Yi}}]{salim2007}
{Salim}, S., {et~al.} 2007, \apjs, 173, 267

\bibitem[{{Salom{\'e}} \& {Combes}(2003)}]{salome03}
{Salom{\'e}}, P., \& {Combes}, F. 2003, \aap, 412, 657

\bibitem[{{Schindler} {et~al.}(2001){Schindler}, {Castillo-Morales}, {De
  Filippis}, {Schwope}, \& {Wambsganss}}]{schindler01}
{Schindler}, S., {Castillo-Morales}, A., {De Filippis}, E., {Schwope}, A., \&
  {Wambsganss}, J. 2001, \aap, 376, L27

\bibitem[{{Schwope} {et~al.}(2000){Schwope}, {Hasinger}, {Lehmann}, {Schwarz},
  {Brunner}, {Neizvestny}, {Ugryumov}, {Balega}, {Tr{\"u}mper}, \&
  {Voges}}]{rbs2}
{Schwope}, A., {et~al.} 2000, Astronomische Nachrichten, 321, 1

\bibitem[{{Seaquist}(1993)}]{1993RPPh...56.1145S}
{Seaquist}, E.~R. 1993, Reports on Progress in Physics, 56, 1145

\bibitem[{{Sijacki} {et~al.}(2007){Sijacki}, {Springel}, {di Matteo}, \&
  {Hernquist}}]{sijacki07}
{Sijacki}, D., {Springel}, V., {di Matteo}, T., \& {Hernquist}, L. 2007,
  \mnras, 380, 877

\bibitem[{{Silk} \& {Rees}(1998)}]{1998A&A...331L...1S}
{Silk}, J., \& {Rees}, M.~J. 1998, \aap, 331, L1

\bibitem[{{Soker}(2006)}]{2006NewA...12...38S}
{Soker}, N. 2006, New Astronomy, 12, 38

\bibitem[{{Sun} {et~al.}(2007{\natexlab{a}}){Sun}, {Donahue}, \&
  {Voit}}]{2007ApJ...671..190S}
{Sun}, M., {Donahue}, M., \& {Voit}, G.~M. 2007{\natexlab{a}}, \apj, 671, 190

\bibitem[{{Sun} {et~al.}(2007{\natexlab{b}}){Sun}, {Jones}, {Forman},
  {Vikhlinin}, {Donahue}, \& {Voit}}]{coronae}
{Sun}, M., {Jones}, C., {Forman}, W., {Vikhlinin}, A., {Donahue}, M., \&
  {Voit}, M. 2007{\natexlab{b}}, \apj, 657, 197

\bibitem[{{Tremaine} {et~al.}(2002){Tremaine}, {Gebhardt}, {Bender}, {Bower},
  {Dressler}, {Faber}, {Filippenko}, {Green}, {Grillmair}, {Ho}, {Kormendy},
  {Lauer}, {Magorrian}, {Pinkney}, \& {Richstone}}]{2002ApJ...574..740T}
{Tremaine}, S., {et~al.} 2002, \apj, 574, 740

\bibitem[{{Vikhlinin} {et~al.}(2005){Vikhlinin}, {Markevitch}, {Murray},
  {Jones}, {Forman}, \& {Van Speybroeck}}]{2005ApJ...628..655V}
{Vikhlinin}, A., {Markevitch}, M., {Murray}, S.~S., {Jones}, C., {Forman}, W.,
  \& {Van Speybroeck}, L. 2005, \apj, 628, 655

\bibitem[{{Wise} {et~al.}(2007){Wise}, {McNamara}, {Nulsen}, {Houck}, \&
  {David}}]{hydraa}
{Wise}, M.~W., {McNamara}, B.~R., {Nulsen}, P.~E.~J., {Houck}, J.~C., \&
  {David}, L.~P. 2007, \apj, 659, 1153

\end{thebibliography}

\clearpage
\begin{deluxetable}{ccccccccccc}
  \tablecolumns{11}
  \tablewidth{0pc}
  \tabletypesize{\small}
  \tablecaption{Cavity Properties.\label{tab:cavities}}
  \tablehead{
    \colhead{Config.} & \colhead{ID} & \colhead{$a$} & \colhead{$b$} & \colhead{$\rlos$} & \colhead{$D$} & \colhead{\tsonic} & \colhead{\tbuoy} & \colhead{\trefill} & \colhead{\ecav} & \colhead{\pcav}\\
    \colhead{-} & \colhead{-} & \colhead{kpc} & \colhead{kpc} & \colhead{kpc} & \colhead{kpc} & \colhead{Myr} & \colhead{Myr} & \colhead{Myr} & \colhead{$10^{60}$ erg} & \colhead{$10^{45}$ erg s$^{-1}$}\\
    \colhead{(1)} & \colhead{(2)} & \colhead{(3)} & \colhead{(4)} & \colhead{(5)} & \colhead{(6)} & \colhead{(7)} & \colhead{(8)} & \colhead{(9)} & \colhead{(10)} & \colhead{(11)}}
  \startdata
  1 & E1      & $17.3 \pm 1.7$ & $10.7 \pm 1.1$ & $13.6 \pm 1.4$ & $23.2 \pm 2.3$ & $20.1 \pm 3.1$ & $28.1 \pm 4.4$ & $70.3 \pm  9.9$ & $1.51 \pm 0.35$ & $1.70 \pm 0.48$\\
  1a & \nodata & \nodata        & \nodata        & $23.4 \pm 2.3$ & \nodata        & \nodata        & \nodata        & $76.9 \pm 10.9$ & $2.59 \pm 0.61$ & $2.92 \pm 0.83$\\
  1 & W1      & \nodata        & \nodata        & $13.6 \pm 1.4$ & $25.9 \pm 2.5$ & $20.3 \pm 4.1$ & $33.3 \pm 5.3$ & $74.4 \pm 10.5$ & $1.72 \pm 0.47$ & $1.64 \pm 0.52$\\
  1a & \nodata & \nodata        & \nodata        & $26.0 \pm 2.6$ & \nodata        & \nodata        & \nodata        & $82.8 \pm 11.7$ & $3.29 \pm 0.89$ & $3.13 \pm 0.99$
  \enddata
  \tablecomments{
    Col. (1) Cavity configuration;
    Col. (2) Cavity identification;
    Col. (3) Semi-major axis;
    Col. (4) Semi-minor axis;
    Col. (5) Line-of-sight cavity radius;
    Col. (6) Distance from central AGN;
    Col. (7) Sound speed age;
    Col. (8) Buoyancy age;
    Col. (9) Refill age;
    Col. (10) Cavity energy;
    Col. (11) Cavity power using \tbuoy.}
\end{deluxetable}

\begin{deluxetable}{ccccc}
  \tablecolumns{5}
  \tablewidth{0pc}
  \tablecaption{Energetic and Mass Accretion Totals.\label{tab:totals}}
  \tablehead{\colhead{Row} & \colhead{Param.} & \colhead{Units} & \colhead{Config-1} & \colhead{Config-1a}}
  \startdata
  (1) & \ecav  & $10^{60}$ erg  & $3.23 \pm 1.16$ & $5.88 \pm 2.11$\\
  (2) & \pcav  & $10^{45}$ \lum & $3.34 \pm 1.41$ & $6.05 \pm 2.56$\\
  (3) & \macc  & $10^7$ \msol   & $1.81 \pm 0.65$ & $3.29 \pm 1.18$\\
  (4) & \dmacc & \msolpy        & $0.59 \pm 0.25$ & $1.07 \pm 0.45$\\
  (5) & \ddmbh & $10^7$ \msol   & $1.63 \pm 0.59$ & $2.96 \pm 1.06$\\
  (6) & \dmbh  & \msolpy        & $0.53 \pm 0.23$ & $0.96 \pm 0.41$\\
  (7) & \dme   & --             & $0.02$          & $0.03$\\
  (8) & \dmb   & --             & $2130$          & $3860$
  \enddata
  \tablecomments{
    Row (1) Cavity energy;
    Row (2) Cavity power;
    Row (3) Mass accreted;
    Row (4) Mass accretion rate;
    Row (5) Black hole mass increase;
    Row (6) Black hole mass growth rate;
    Row (7) Eddington ratio;
    Row (8) Bondi ratio.}
\end{deluxetable}

\begin{deluxetable}{lcccccccccccc}
  \tablecolumns{13}
  \tablewidth{0pc}
  \tabletypesize{\footnotesize}
  \tablecaption{Nuclear X-ray Point Source Spectral Models.\label{tab:agn}}
  \tablehead{
    \colhead{Absorber} & \colhead{\nhabs} & \colhead{$\Gamma_{\rm{pl}}$} & \colhead{$\eta_{\rm{pl}}$} & \colhead{$E_{\rm{ga}}$} & \colhead{$\sigma_{\rm{ga}}$} & \colhead{$\eta_{\rm{ga}}$} & \colhead{Param.} & \colhead{$L_{0.7-7.0}$} & \colhead{\lbol} & \colhead{\chisq} & \colhead{DOF} & \colhead{Goodness}\\
    \colhead{-} & \colhead{$10^{22}~\pcmsq$} & \colhead{-} & \colhead{$10^{-5} \dagger$} & \colhead{keV} & \colhead{eV} & \colhead{$10^{-6} \ddagger$} & \colhead{-} & \colhead{$10^{44}~\lum$} & \colhead{$10^{44}~\lum$} & \colhead{-} & \colhead{-} & \colhead{-}\\
    \colhead{(1)} & \colhead{(2)} & \colhead{(3)} & \colhead{(4)} & \colhead{(5)} & \colhead{(6)} & \colhead{(7)} & \colhead{(8)} & \colhead{(9)} & \colhead{(10)} & \colhead{(11)} & \colhead{(12)} & \colhead{(13)}}
  \startdata
  None          & \nodata             & $0.1^{+0.3}_{-0.3}$ & $0.3^{+0.1}_{-0.1}$  & [2.4, 3.4] & [63, 119] & [8.4, 14.9] & \nodata                & $0.69^{+0.11}_{-0.22}$ & $49.9^{+18.1}_{-19.0}$ & 1.88 & 61 & 56\%\\
  Neutral$^a$   & $4.2^{+1.9}_{-1.3}$ & $1.5^{+0.4}_{-0.3}$ & $6.6^{+8.1}_{-3.6}$  & [1.8, 3.0] & [31, 58] & [0.6, 1.8]  & 0.354                  & $0.68^{+0.12}_{-0.24}$ & $5.65^{+2.11}_{-2.50}$ & 1.17 & 60 & 29\%\\
  Warm$^b$      & $3.3^{+1.4}_{-1.6}$ & $1.9^{+0.2}_{-0.2}$ & $16.2^{+0.4}_{-0.5}$ & [1.8, 2.9] & [57, 44] & [0.9, 1.5]  & $0.97^{+0.03}_{-0.03}$ & $0.70^{+0.18}_{-0.26}$ & $3.46^{+1.10}_{-0.95}$ & 1.01 & 59 & 13\%\\
  Power-law$^c$ & 0.5--7.5            & $2.1^{+0.5}_{-0.3}$ & $22.1^{+2.7}_{-1.1}$ & [1.8, 3.0] & [54, 35] & [0.9, 1.4]  & $0.63^{+0.34}_{-0.31}$ & $0.71^{+1.48}_{-1.32}$ & $2.21^{+0.45}_{-0.30}$ & 1.00 & 58 & $< 1$\%
  \vspace{0.5mm}
  \enddata
  \tablecomments{
    For all models, $\nhgal = 2.28 \times 10^{20} ~\pcmsq$. 
    Col. (1) \xspec\ absorber models: ($a$) is \textsc{zwabs}, ($b$) is \textsc{pcfabs}, ($c$) is \textsc{pwab};
    Col. (2) Absorbing column density;
    Col. (3) Power-law index;
    Col. (4) Power-law normalization with units ($\dagger$) ph keV$^{-1}$ cm$^{-2}$ s$^{-1}$ at 1 keV;
    Col. (5) Gaussian central energies;
    Col. (6) Gaussian widths;
    Col. (7) Gaussian normalizations with units ($\ddagger$) ph cm$^{-2}$ s$^{-1}$;
    Col. (8) Model-dependent parameter: ($a$) absorber redshift, ($b$) absorber covering fraction, ($c$) absorber power law index of covering fraction;
    Col. (9) Model 0.7-7.0 keV luminosity;
    Col. (10) Unabsorbed model bolometric (0.01-100.0 keV) luminosity;
    Col. (11) Reduced \chisq\ of best-fit model;
    Col. (12) Model degrees of freedom;
    Col. (13) Percent of 10,000 Monte Carlo realizations with \chisq\ less than best-fit \chisq.
}
\end{deluxetable}

\begin{deluxetable}{ccccc}
  \tablecolumns{5}
  \tablewidth{0pc}
  \tablecaption{BCG Star Formation Rates.\label{tab:sfr}}
  \tablehead{
    \colhead{Source} & \colhead{ID} & \colhead{$\xi$ [Ref.]} & \colhead{$L$} & \colhead{$\psi$}\\
    \colhead{-} & \colhead{-} & \colhead{(\msolpy)/(\lum ~\phz)} & \colhead{\lum ~\phz} & \colhead{\msolpy}\\
    \colhead{(1)} & \colhead{(2)} & \colhead{(3)} & \colhead{(4)} & \colhead{(5)}}
  \startdata
  \galex\            & NUV             & $1.4 \times 10^{-28}$ [1] & $2.5 ~(\pm 0.9) \times 10^{28}$ & $3.5 \pm 1.3$\\
  \xom\              & UVW1            & $1.1 \times 10^{-28}$ [2] & $6.2 ~(\pm 1.9) \times 10^{28}$ & $6.9 \pm 2.2$\\
  \galex\            & FUV             & $1.1 \times 10^{-28}$ [2] & $8.2 ~(\pm 2.0) \times 10^{28}$ & $9.0 \pm 2.3$\\
  \galex\            & FUV             & $1.4 \times 10^{-28}$ [1] & $8.2 ~(\pm 2.0) \times 10^{28}$ & $11 \pm 3$\\
  \xom\              & UVM2            & $1.1 \times 10^{-28}$ [2] & $< 5.0 \times 10^{29}$          & $< 55$
  \enddata
  \tablecomments{A dagger ($\dagger$) indicates the removal of
    Hz$^{-1}$ from the units of $\xi$ \& $L$.  Col. (1) Source of
    measurement; Col. (2) Diagnostic identification; Col. (3)
    Conversion coefficient and references: [1] \citet{kennicutt2}, [2]
    \citet{salim2007}; Col. (4) Luminosity; Col. (5) Star formation
    rate.}
\end{deluxetable}

\end{document}